\documentclass[useAMS,usenatbib,twocolumn]{mnras}

\usepackage{amsmath}
\usepackage{graphicx}
\usepackage{dcolumn}
\usepackage{multicol}
\usepackage{bm}
\usepackage{epsfig}
\usepackage{amssymb,latexsym,mathrsfs}
\usepackage{graphicx}
\usepackage{enumitem}
\usepackage{color}
\usepackage{hyperref}

\usepackage[normalem]{ulem}
\usepackage{float}
\usepackage{natbib}
\usepackage[utf8]{inputenc}
\usepackage[T1]{fontenc}

\makeatletter
\def\hlinewd#1{%
\noalign{\ifnum0=`}\fi\hrule \@height #1 \futurelet
\reserved@a\@xhline}
\makeatother

\usepackage[dvipsnames]{xcolor}

\hypersetup{
    colorlinks=true,
    linkcolor=red,
    citecolor=blue,
} 

\def\apj{{ ApJ}}
\def\apjl{{ ApJL}}
\def\apjs{{ ApJS}}

\def\aap{{ A\&A}}

\def\mnras{{ MNRAS}}

\def\nat {{ Nature}}

\def\prd{{ Phys. Rev. D}}

\newcommand{\be}{\begin{equation}}
\newcommand{\ee}{\end{equation}}
\newcommand{\bs}{\begin{split}} 
\newcommand{\bea}{\begin{eqnarray}}
\newcommand{\eea}{\end{eqnarray}}

\title[High-energy neutrinos from protomagnetar jets]{High-energy neutrino emission from magnetized jets of rapidly rotating protomagnetars} %Highly magnetized protomagnetar jets: propagation \& neutrino emission

\author[Bhattacharya et al.]{
Mukul Bhattacharya$^{1}$\thanks{mmb5946@psu.edu}, Jose A. Carpio$^{1}$, Kohta Murase$^{1,2,3}$, Shunsaku Horiuchi$^{4,5}$\\
${}^1$Department of Physics; Department of Astronomy \& Astrophysics; Center for Multimessenger Astrophysics,\\
Institute for Gravitation and the Cosmos, The Pennsylvania State University, University Park, PA 16802, USA\\
${}^2$School of Natural Sciences, Institute for Advanced Study, Princeton, New Jersey 08540, USA\\
${}^3$Center for Gravitational Physics and Quantum Information, Yukawa Institute for Theoretical Physics, Kyoto University, Kyoto,\\ Kyoto 606-8502, Japan\\
${}^4$Center for Neutrino Physics, Department of Physics, Virginia Tech, Blacksburg, VA 24061, USA\\
${}^5$Kavli IPMU (WPI), UTIAS, The University of Tokyo, Kashiwa, Chiba 277-8583, Japan 
} 

\begin{document}

\date{Accepted . Received ; in original form }

\pagerange{\pageref{firstpage}--\pageref{lastpage}} \pubyear{2023}

\maketitle

\label{firstpage}

\begin{abstract} 
Relativistic jets originating from protomagnetar central engines can lead to long duration gamma-ray bursts (GRBs) and are considered potential sources of ultrahigh-energy cosmic rays and secondary neutrinos. We explore the propagation of such jets through a broad range of progenitors, from stars which have shed their envelopes to supergiants which have not. 
We use a semi-analytical spindown model for the strongly magnetized and rapidly rotating protoneutron star (PNS) to investigate the role of central engine properties such as the surface dipole field strength, initial rotation period, and jet opening angle on the interactions and dynamical evolution of the jet-cocoon system. 
With this model, we determine the properties of the relativistic jet, the mildly-relativistic cocoon, and the collimation shock in terms of system parameters such as the time-dependent jet luminosity, injection angle and density profile of the stellar medium. 
We also analyse the criteria for a successful jet breakout, the maximum energy that can be deposited into the cocoon by the relativistic jet, and structural stability of the magnetized outflow relative to local instabilities. Lastly, we compute the high-energy neutrino emission as these magnetized outflows burrow through their progenitors. Precursor neutrinos from successful GRB jets are unlikely to be detected by IceCube, which is consistent with the results of previous works. On the other hand, we find high-energy neutrinos may be produced for extended progenitors like blue and red supergiants, and we estimate the detectability of neutrinos with next-generation detectors such as IceCube-Gen2. 
\end{abstract}


\section{Introduction}
Relativistic jets are commonly inferred in a wide variety of compact astrophysical systems and can be powered by, e.g., magnetic extraction of neutron star (NS) rotational energy or accretion of infalling mass onto a black hole (BH). 
The extracted energy is transported outward as the Poynting flux, and converted to kinetic energy flux either gradually \citep{HN1989,Lyubarsky2009} or implusively \citep{Granot2011,Lyutikov2011}, and magnetic dissipation may be important for efficient acceleration \citep{Usov1992,Drenkhahn2002}. 
Ultra-relativistic jets that break out from their dense host medium can potentially power gamma-ray bursts (GRBs, see e.g., \citealt{Meszaros2006}, for a review), and the association of long duration GRBs with core-collapse supernovae (SNe) further indicates their production in stellar core-collapse. 
Sufficiently low-power jets, including low-luminosity (LL) GRBs and ultra-long (UL) GRBs, may arise if the relativistic jet gets smothered by extended stellar material or large stellar progenitors \citep{Campana2006,Toma2007,IC2016}. 
Such jets have been of special interest as they exhibit intermediate properties between GRBs and transrelativistic supernovae \citep{Soderberg2006}, thereby potentially providing a unified picture for the GRB-SNe connection \citep{Margutti2013,Margutti2014,Nakar:2015tma}.

Protomagnetars have been the subject of great interest as the central engine of GRBs \citep{Thompson2004,Metzger2011a}. 
Observations of flares and extended emission suggest that some central engines can be active for a long time (e.g., \citealt{Romano:2006rd,Dai:2006hj,Metzger:2007cd,Zhang:2013coa}), which could be powered by the protomagnetar. It is known that pulsar winds are highly relativistic, and the tenuous nebular drives the ejecta, that powers super-luminous SNe (SLSNe) and aid SNe Ibc \citep{Kazumi2016,Margalit:2017oxz,Margalit:2018bje}. The magnetar outflow is also an interesting site for nucleosynthesis of heavy elements \citep[e.g.,][]{Metzger2011b,Horiuchi2012,MBh2021,Ekanger2022}. 

An unavoidable consideration in all these systems is the jet interaction with the stellar and extended external media. This interaction determines the jet system dynamics and, e.g., affects the jet velocity, collimation and breakout criterion. There has been significant progress in understanding this physics. Many analytical (e.g., \citealt{BR1974,BC1989,MW2001,Matzner2003,LB2005,Bromberg2011}) and numerical (e.g., \citealt{Aloy2000,Zhang2004,Lazzati2009,MA2009,Nagakura2011,MI2013,Harrison2018,Gottlieb2019}) efforts have been made to investigate the propagation of hydrodynamic jets. In recent years, some numerical works have included the effect of magnetic fields to study how they alter the hydrodynamic picture (e.g., \citealt{UM2007,Bucci2009,LB2013,BT2016,Bromberg2019,Matsumoto2021,Nathanail2021}). 
Magnetized jets tend to have narrower cross-section compared to hydrodynamic jets and propagate at relativistic velocities with shorter breakout times \citep{Bromberg2012,Bromberg2015}.

Relativistic outflows can be collimated by oblique shocks that form inside the outflow close to the jet base and converge on the jet axis (\citealt{Bromberg2011}). Jet collimation reduces the jet-head cross section, thereby accelerating its propagation through the surrounding medium. 
Observations indicate that the opening angle of long duration GRBs can be distributed over several to tens of degrees (see e.g., \citealt{Fong2012}). Interactions of jets with the surrounding dense medium promotes the growth of local instabilities along the jet-cocoon boundary. 
If these instabilities grow to large amplitudes, they can lead to substantial entrainment of baryons into the jet which can alter the jet dynamics and its emission properties \citep{Aloy2000,Mac2001,Gottlieb2019,MM2019}. Strong mixing of jet material with the cocoon prior to breakout leads to heavy baryon loading especially after the collimation point (see e.g., \citealt{PIM2021}). The degree of mixing is strongly influenced by the jet power, injection angle and density of the stellar medium \citep[e.g.,][]{Gottlieb2020a}: high-power jets with small injection angle propagating in low density medium develop faster moving jet-head and show a smaller degree of mixing.

Due to the challenges in probing jets embedded deep inside a star, neutrinos have been of interest as one of the possible messengers \citep[e.g.,][]{MW2001,Razzaque:2002kb,Razzaque:2004yv,Ando:2005xi,Horiuchi:2007xi}. 
The signatures are often called precursor neutrinos for successful jets and orphan neutrinos for choked jets. However, neutrino production is likely to be inefficient in radiation-dominated jets as shown in \cite{MuraseIoka2013} (see their Fig.~3), where high-luminosity GRB jets propagating in a Wolf-Rayet (WR) star were essentially excluded as precursor/orphan neutrino sources. 
This is especially the case for slow jets including collimated jets where the jet is radiation dominated\footnote{Note that \citet{MuraseIoka2013} provide generic criteria for a given shock radius and the Lorentz factor at this radius, which should be determined by simulations or data. Importantly, in contrast to previous works \citep{MW2001,Razzaque:2003uw,Razzaque:2004yv,Ando:2005xi}, they largely exclude cosmic-ray acceleration in typical GRB jets, especially after the collimation point, where the jet is radiation dominated and the Lorentz factor is only a few, independent of the maximum attainable Lorentz factor. \citet{Guarini:2022uyp} argue that their results are contrary to \citet{MuraseIoka2013}, which is incorrect. \citet{MuraseIoka2013} actually showed that shock acceleration in slow jets in a WR star is unlikely, while the allowed parameter space essentially corresponds to that unexplored by \citet{Guarini:2022uyp}.}. 
Except for special situations of subshock formation \citep{GG2021}, the radiation constraint is so severe that fast jet acceleration and/or extended material are necessary \citep{MuraseIoka2013}. Jets inside much more extended material, e.g., blue supergiants (BSGs), red supergiants (RSGs), and circumstellar material, have been considered in the literature \citep{Senno:2015tsn, He:2018lwb, Guarini:2022uyp,Soker2021}. 
For example, uncollimated jets can be expected for low-luminosity GRBs \citep{Senno:2015tsn} and jets from supermassive black hole mergers \citep{Yuan:2021jjt}. In the protomagnetar model, the outflow can achieve large Lorentz factors especially at late times \citep{Metzger2011a}, which we focus upon in this work.

To this end, we explore a self-consistent semi-analytical description for magnetized outflows that arise from PNS central engines, as they propagate inside and across stellar progenitors including WRs, BSGs and RSGs (see e.g., \citealt{WW1995,WH2006}). 
The dynamics of the magnetized outflow is determined by its time-dependent luminosity and magnetization which are obtained from the PNS spin-down evolution \citep{Metzger2011a}. It should be noted that even state-of-art magnetohydrodynamic simulations do not fully explore the effects of central engine spin-down and continued neutrino-driven mass loss on the magnetized outflow in a time-dependent manner, especially at late times where the magnetization is expected to be very large. The jet is typically Poyting-dominated even if it acquires a large Lorentz factor, so we consider magnetic reconnection as a viable mechanism for particle acceleration \citep{Hoshino2013,Guo_2016,Xiao2017,Ball_2018}. 

The focus of this work is to model the evolution of the magnetized jet-cocoon system; study its dependence on the physical parameters such as jet luminosity, injection angle and density profile of the external medium; and compute its high-energy neutrino emission and detectability at IceCube-Gen2. This paper is organised as follows. In Section~\ref{Sec2}, we provide a brief overview of the PNS spin-down evolution and the input parameters used in our semi-analytical study. 
In Section~\ref{Sec3}, we describe the analytical model used to study the propagation of magnetized jets and potential collimation from interaction with the surrounding stellar medium.  
In Section~\ref{Sec4}, we examine the energy requirement for a successful jet breakout and analyse the structural stability of these jets relative to local magnetic instabilities. We compute the time-dependent energy spectra of high-energy neutrinos and estimate their detectability in Section~\ref{Sec5}. 
Finally, we discuss the main implications of our results in Section~\ref{Sec6} and summarise our conclusions in Section~\ref{Sec7}. 
The symbols used in this work are listed along with their physical description in Table~\ref{Table1}.

\section{Description of physical system}
\label{Sec2}
Here we describe the spin-down evolution of rapidly rotating and strongly magnetized PNS, and its effect on the neutrino-driven winds that eventually power the relativistic jet. We estimate the time-dependent jet luminosity from outflow magnetization and mass-loss rate for system parameters such as the dipole field strength, initial rotation period, jet injection angle and density profile of the stellar medium.

\subsection{Protomagnetar central engine}\label{S2.1}
We adopt the analytical prescription of \citet{QW1996} and \citet{Metzger2011a} to model the properties of neutrino-driven winds from PNS. 
The mass loss rate from PNS surface due to neutrino-heated wind is given by
\begin{equation}\label{Mdot}
    \Dot{M}=(5\times10^{-5}~\textrm{M}_\odot~\textrm{s}^{-1})f_{\rm open}f_{\rm cent}C_{\rm es}^{5/3}L_{\nu,52}^{5/3}\epsilon_{\nu,10}^{10/3}R_{10}^{5/3}M_{1.4}^{-2}
\end{equation}
where $f_{\rm open}$ is the fraction of PNS surface threaded by open magnetic field lines, $f_{\rm cent}$ accounts for the magnetocentrifugal enhancement to $\dot{M}$,  
$C_{\rm es}$ is heating correction factor for inelastic neutrino-electron scatterings, $L_{\nu}=L_{\nu,52}\times10^{52}\, {\rm erg\, s^{-1}}$ is the neutrino luminosity, $\epsilon_{\nu}=\epsilon_{\nu,10}\times10\, {\rm MeV}$ is the mean neutrino energy, $R_{\rm NS}=R_{\rm NS, 10}\times10\, {\rm km}$ is the PNS radius and $M_{\rm NS}=M_{\rm NS, 1.4}\times1.4\, M_{\odot}$ is the PNS mass. 
We use the same definition for time-dependent $\dot{M}$ correction factors, $f_{\rm open}$ and $f_{\rm cent}$, as in \citet{Metzger2011a}. Following \citet{Pons1999}, we adopt the dynamical neutrino quantities for a PNS with $M_{\rm NS}=1.4\, M_{\odot}$. 
As \citet{Pons1999} do not account for effects of rapid rotation on PNS cooling, we follow \citet{Metzger2011a} and include a stretch factor $\eta_s=3$ to appropriately model the cooling as: $L_{\nu}\rightarrow L_\nu|_{\Omega=0}~\eta_s^{-1},\ t\rightarrow t|_{\Omega=0}~\eta_s,\ \epsilon_{\nu}\rightarrow \epsilon_{\nu}|_{\Omega=0}~\eta_s^{-1/4}$ where the subscript $\Omega=0$ represents non-rotating PNS quantities. 

\begin{figure} 
\includegraphics[width=1.0\columnwidth]{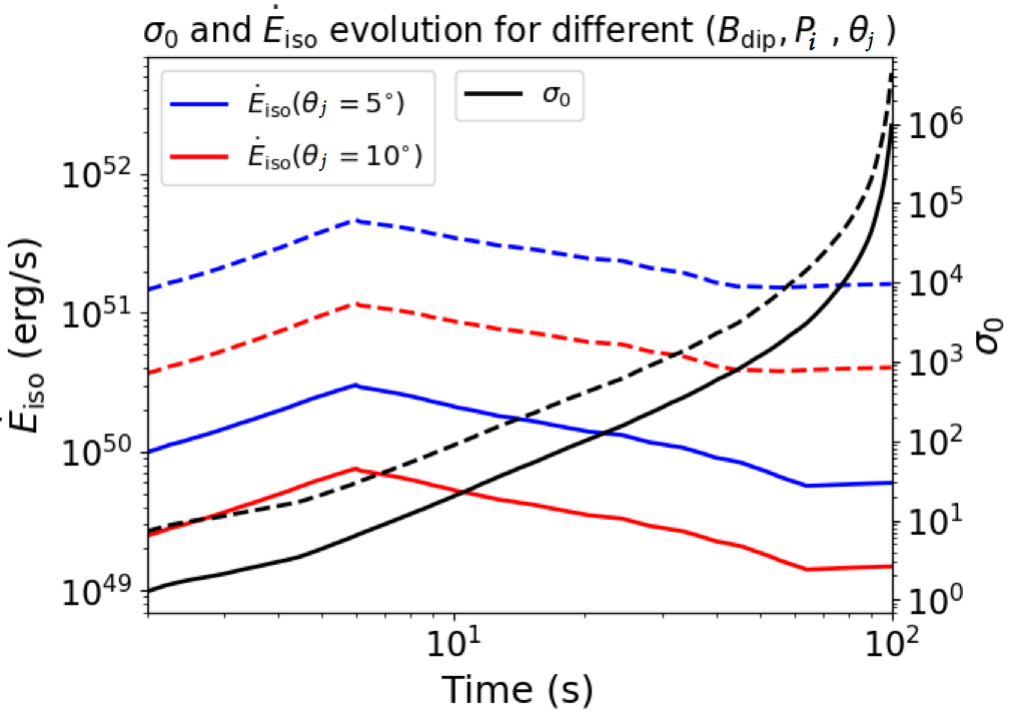} 
\vspace{-0.5cm}
\caption{Time variation of jet magnetization $\sigma_0$ and isotropic luminosity $\dot{E}_{\rm iso}=\dot{E}_{\rm tot}/(\theta_j^2/2)$  
are shown for magnetic obliquity angle $\chi=\pi/2$. The wind properties are evaluated for configurations with $(B_{\rm dip},P_i)=(10^{15}\, {\rm G},2\, {\rm ms})$ [solid curves] and $(3\times10^{15}\, {\rm G},1.5\, {\rm ms})$ [dashed curves]. For both these cases, the blue curves correspond to an initial jet opening angle $\theta_j=5^{\circ}$ whereas the red curves are those for $\theta_j=10^{\circ}$.} 
\label{sigma0_Edot_t}
\end{figure}

The outflow magnetization equals the Lorentz factor that the jet attains once its magnetic energy is fully converted into the bulk kinetic energy, and is given by $\sigma_0=\phi_B^2\Omega^2/\Dot{M}c^3$, where $\phi_B=(f_{\rm open}/4\pi)B_{\rm dip}R_{\rm NS}^2$ is the magnetic flux due to a rotating dipole field with magnitude $B_{\rm dip}$ and $\Omega$ is the PNS angular velocity. 
The radial distance $r$ from the central engine is obtained by solving the equation (see also, e.g. \citealt{Drenkhahn2002})
\begin{equation}
\label{Gamma_j}
\beta_j\Gamma_j=
\begin{cases} 
      \sigma_0(r/R_{\textrm{mag}})^{1/3}, & r\leq R_{\textrm{mag}} \\
      \sigma_0, & r>R_{\textrm{mag}} 
\end{cases}
\end{equation}
where $\Gamma_j = (1-\beta_j^2)^{-1/2}$ is the jet Lorentz factor and the magnetic dissipation radius is 
\begin{equation}
\label{R_mag}
    R_{\rm mag}=(5\times10^{12}\, {\rm cm})\left(\frac{\sigma_0}{10^2}\right)^2\left(\frac{P}{\rm ms}\right)\left(\frac{\epsilon}{0.01}\right)^{-1}.
\end{equation}
Here, $\epsilon = v_r/c \approx 0.01$ parametrizes the reconnection velocity $v_r$. GRB emission is initiated when 
$r \sim R_{\rm mag}$, and is typically powered by the dissipation of jet's Poynting flux near the photosphere. 
From Equation (\ref{Gamma_j}), the minimum outflow Lorentz factor
is obtained at $R=R_L$, where $R_L=c/\Omega$ is the light cylinder radius.

The pulsar wind would not be highly collimated at the light cylinder, and the wind in the open field zone creates a tenuous bubble or cavity. The wind, which pushes the ejecta including the cocoon material, is significantly decelerated after the wind termination shock, forming a hot magnetized bubble. The evolution of the cavity and nebula depends on the spin-down power of the PNS and the ejecta \citep[e.g.,][]{CF1992,Bucci2009,Kotera2013}. Magnetic dissipation inside the tenuous wind bubble with radius $R_w$ has been considered in the context of magnetar models for broadline SNe Ibc, SLSNe and rapidly rising optical transients \citep{Kashi2016,Hotokez2017,Margalit:2017oxz,Margalit:2018bje}. For the typical magnetic fields and spin periods for the PNS considered here, we obtain $R_w \sim 10^9-10^{10}\, {\rm cm}$ at $\sim {\rm few}\ 10\, {\rm s}$ \citep[e.g.,][]{Kotera2013,MKP2016}. The mixing with the external cocoon material may be prohibited for jets inside the bubble, and the jet can attain large Lorentz factor before interacting with the stellar material.

The kinetic wind luminosity is $\dot{E}_{\rm kin} = (\Gamma_{j}-1)\dot{M}c^2$, where $\Gamma_{j}$ is determined by equation~(\ref{Gamma_j}). Relativistic outflows achieve $\Gamma_j \approx \sigma_0^{1/3}$ at $r = R_{\rm mag}$ and therefore $\dot{E}_{\rm kin}$ is independent of $\dot{M}$. As magnetic power is determined by the Poynting flux $\phi_B$, it is given as $\dot{E}_{\rm mag}=(2/3)\dot{M}c^2 \sigma_0$ for relativistic outflows \citep{Metzger2011a}. For such outflows, most of the wind power resides in Poynting flux as $\dot{E}_{\rm mag}/\dot{E}_{\rm kin} \sim \sigma_0^{2/3} \gg 1$. 
The rapidly rotating PNS gradually loses its angular momentum $J=(2/5)M_{\rm NS}R_{\rm NS}^2 \Omega$ to the wind at a rate $\dot{J} = -\dot{E}_{\rm tot}/\Omega$. 
PNS spin-down is solved from its mass $M_{\rm NS}$, surface dipole field $B_{\rm dip}$, initial rotation period $P_i = 2\pi/\Omega_i$ and magnetic obliquity angle $\chi$. As PNS continues to contract for first few seconds post core-collapse, $\Omega_i$ and $B_{\rm dip}$ are defined as the maximum values that are achieved when the PNS contracts with conserved angular momentum $J \propto R_{\rm NS}^2 M_{\rm NS} \Omega$ and conserved magnetic flux $\phi_B \propto B_{\rm dip}R_{\rm NS}^2$.

Figure~\ref{sigma0_Edot_t} shows the time evolution of $\sigma_0$ and $\dot{E}_{\rm iso}=\dot{E}_{\rm tot}/(\theta_j^2/2)$ for magnetized jets with fixed obliquity angle $\chi=\pi/2$. The outflow properties are shown for two central engine $(B_{\rm dip},P_i)$ configurations and for jet opening angle $\theta_j=5^{\circ},10^{\circ}$. Irrespective of the values of $B_{\rm dip}$, $P_i$ and $\theta_j$, the initially non-relativistic outflow evolves to a relativistic state with $\sigma_0 \gtrsim 1$ over the first few seconds. While the jet magnetization does not depend on $\theta_j$, $\sigma_0 \propto B_{\rm dip}^2 \Omega^2$ tends to be larger for PNS with strong field and rapid rotation. 
The wind isotropic luminosity $\dot{E}_{\rm iso}$ increases until $t \sim 5-10\, {\rm s}$ as both $B_{\rm dip}$ and $\Omega$ grow due to conservation of $\phi_B$ and $J$, respectively, while the PNS shrinks in radius. Once the contraction stops, $\dot{E}_{\rm iso}$ gradually decreases as the PNS continues to spin down and finally saturates after $t \sim 30-50\, {\rm s}$. As expected, for central engines with similar $B_{\rm dip}$ and $P_i$, jets with smaller opening angle tend to be more energetic. The stronger magnetocentrifugal acceleration in these outflows leads to a larger energy loss rate and therefore $\dot{E}_{\rm iso}$.

\subsection{Initial parameters}\label{S2.3}
Magnetized jet dynamics and interaction with stellar medium is primarily determined by the surface dipole field $B_{\rm dip}$, initial rotation period $P_i$, jet opening angle $\theta_{j}$ and density profile $\rho_{a}(r)$ of the external medium. 
In the spirit of a parameter study, we treat these physical parameters as independent of each other. 
We discuss the effect of each of these physical parameters in more detail below.  

\begin{figure}
\centering
\includegraphics[width=0.96\columnwidth]{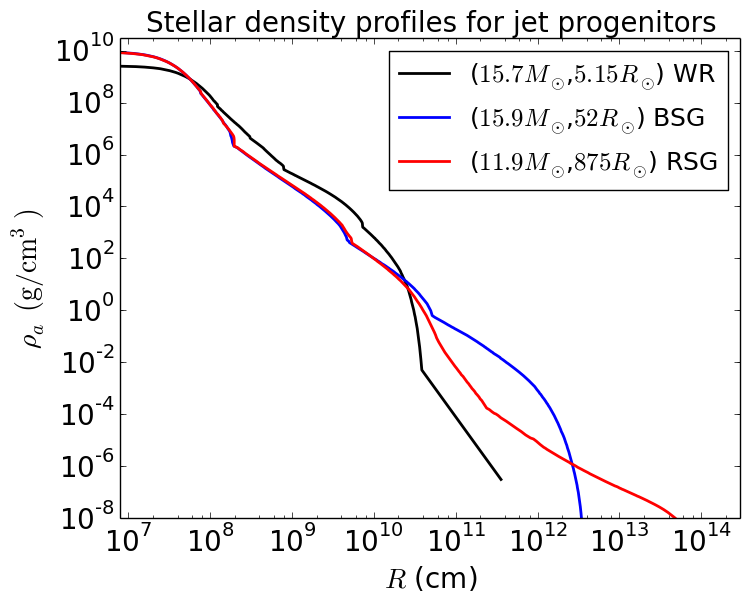}
\vspace{-0.3cm} 
\caption{Stellar density profiles are shown for the three  
progenitor models considered in this work: the ($15.7\, M_{\odot}$, $5.15\, R_{\odot}$) Wolf-Rayet star, the ($15.9\, M_{\odot}$, $52\, R_{\odot}$) blue supergiant with $Z=0.01\, Z_{\odot}$, and the ($11.9\, M_{\odot}$, $875\, R_{\odot}$) red supergiant with $Z=Z_{\odot}$. The evolution of mass density is shown as a function of the radial distance $R$ from the central PNS.  
} 
\label{density_profiles}
\end{figure}

\begin{figure*} 
\includegraphics[width=0.46\textwidth]{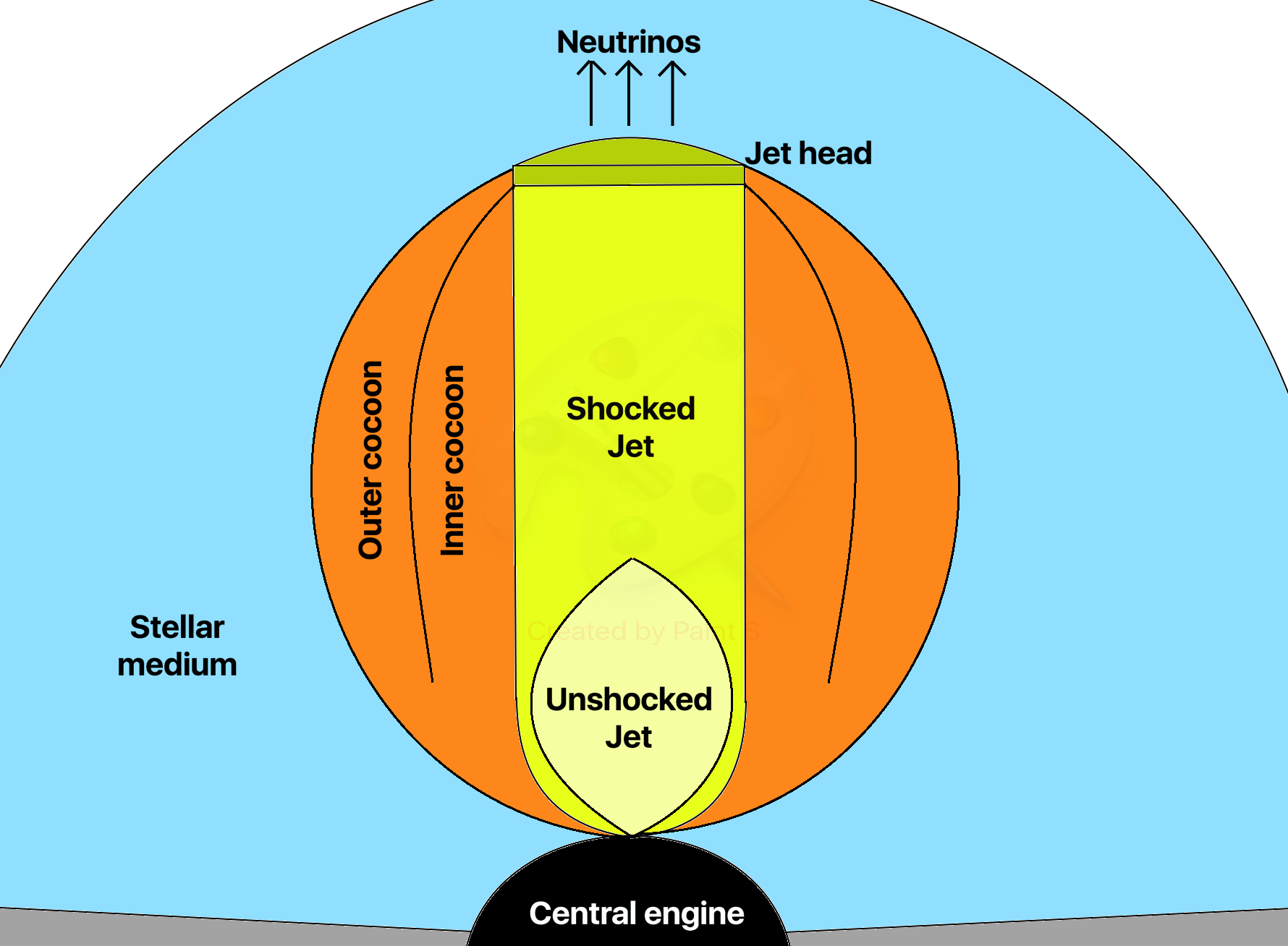} 
\hspace{0.5cm}
\includegraphics[width=0.46\textwidth]{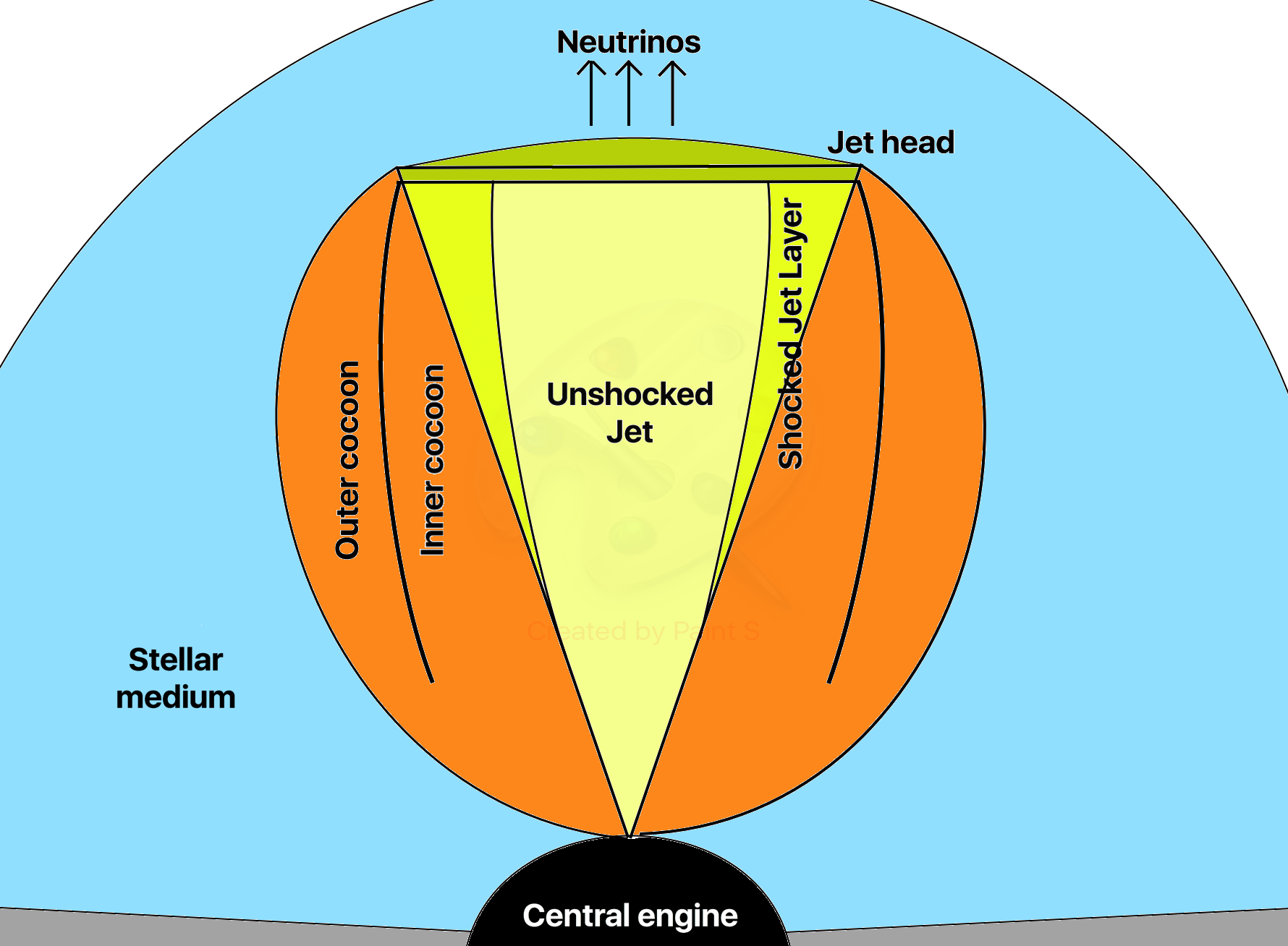}
\caption{A schematic description of the structure of the jet-cocoon system in the two collimation regimes (left panel shows a collimated jet whereas the right panel shows an uncollimated jet), distinguished by the jet luminosity, initial opening angle and density of the surrounding medium.  The five main components are shown: the jet head, the jet divided into shocked and unshocked parts by the CS, the inner and outer regions of the cocoon separated by the contact discontinuity, and the stellar medium. 
} 
\label{schematic}
\end{figure*}

\begin{itemize}[leftmargin=*]
    \begin{item} \textbf{Magnetic field and obliquity:} We consider surface dipole field, 
    $3\times10^{14}\,{\rm G} \lesssim B_{\rm dip} \lesssim 3\times10^{16}\,{\rm G}$. If PNS magnetic energy  
    is roughly equal to its rotational energy, fields 
    up to $\sim 3\times10^{17}\, {\rm G}$ can be achieved. However, stable 
    configurations require the dipole component to be at least 10 times smaller than the total field strength, thereby giving $B_{\rm dip} \lesssim 3\times10^{16}\, {\rm G}$. PNS with dipole field component below $\sim 10^{14}\, {\rm G}$ are generally not considered as protomagnetars. 
    
    The obliquity angle between the magnetic and rotational axes is $0 \lesssim \chi \lesssim \pi/2$. Although both $B_{\rm dip}$ and $\chi$ can affect the mass-loss rate and baryon loading of the outflow \citep{Shibata2011}, here we assume that $\dot{M}$ is primarily determined by neutrino heating as given by equation~(\ref{Mdot}).\\ 
    \end{item}
    \begin{item} \textbf{Initial rotation period:} In this study, we treat the initial spin period ($1\, {\rm ms} \lesssim P_i \lesssim 5\, {\rm ms}$) as an independent parameter. While $P_i \gtrsim 1\, {\rm ms}$ is determined by the allowed range of stable PNS rotational periods \citep{Strobel1999}, the maximum rotation period $P_i \sim 5\,{\rm ms}$ is imposed by the angular momentum loss that is incurred by the rapidly spinning PNS to magnetorotational instability-generated turbulence (see e.g. \citealt{Ott2005}).  
    Rapid rotation leads to higher $\dot{E}_{\rm tot}$ due to larger magnetocentrifugal acceleration, and therefore, an increased feasibility to generate jets that can break out. 
    Enhancement in $\dot{M}$ due to magnetocentrifugal forces is most significant when the PNS is rapidly rotating and has a large magnetic obliquity angle.\\
    \end{item}
    \begin{item} \textbf{Jet initial opening angle:}  
    Jet opening angle affects GRB energy output and its observed rate. Observations of GRB outflows indicate that the initial jet opening angle $\theta_{j}$ for long duration GRBs can vary within a broad range of several to tens of degrees \citep{Fong2012}. As $\theta_{j} \sim 1/\Gamma_{\rm j}$, it is determined by $\sigma_0$ which itself is a function of the central engine parameters $B_{\rm dip}$, $P_i$ and $\chi$ (see equation~\ref{Gamma_j}). In this study, we will assume $\theta_j=5^{\circ}, 10^{\circ}, 20^{\circ}$ as the three representative cases for magnetized jets. Jet collimation through oblique shocks at later times can also impact the opening angle as shocks converge the outflow to jet axis and reduce jet-head cross section, thereby reducing the opening angle.\\ 
    \end{item}
    \begin{item} \textbf{PNS mass:} Larger PNS mass leads to smaller $\sigma_0$, while $\dot{E}_{\rm tot}$ is only marginally elevated (see \citealt{MBh2021}). Furthermore, $\dot{M}$ is also amplified by higher PNS mass. To simplify our model, we fix PNS baryonic mass to $M_{\rm NS}=1.4\,M_{\odot}$ for all calculations and use corresponding neutrino cooling curves from \citet{Pons1999}, corrected for rotation as described in the previous subsection.  \\
    \end{item}
    \begin{item} \textbf{Stellar density profile:}
    The density profile in the stellar envelope of GRB progenitors can be roughly approximated as 
    \begin{equation}
    \rho_{a}(r) = \overline{\rho}\left(\frac{r}{R}\right)^{-\alpha} = \left(\frac{3-\alpha}{4\pi}\right)\frac{M}{R^3}\left(\frac{r}{R}\right)^{-\alpha}
    \end{equation}
    where $\alpha \sim 2-3$ is the density profile index, $\overline{\rho}$ is the average density, $M$ is the mass of the stellar progenitor with radius $R$ \citep{MM1999}. Density is only a function of height $r$, which is a good approximation since opening angles of jet and cocoon are relatively small. 

    We adopt three progenitor scenarios as representative cases of massive stars that have expended all available nuclear fuel just prior to collapse. The first is a rapidly rotating young Wolf-Rayet (WR) star, which 
    represent the final evolutionary stage of the most massive stars that have depleted their hydrogen/helium envelope due to either winds or binary interactions. Due to their association with observed GRBs, they are strongly motivated for powering jets. Since they lack an envelope, WR stars are generally compact. Second, we consider massive stars which have retained their envelopes until core-collapse: blue supergiants (BSGs) and red supergiants (RSGs). These stars have significantly larger radii ($R \gtrsim 50\, R_{\odot}$) than WR stars (see e.g., \citealt{Matzner2003}). No supergiant has yet been found as GRB progenitor, but jet-like outflows are harder to detect in supergiants in contrast to their WR counterparts. Jet launch should be associated with properties such as the core size, magnetic field, or rotation rate, and the connection to the envelope is not clear. Thus, we also consider supergiants as hosts of jets.  
    For our calculations, we consider three stellar density profiles (see e.g., \citealt{WW1995,WH2006}): a $15.7\, M_{\odot}$ and $5.15\, R_{\odot}$ WR star, a $15.9\, M_{\odot}$ and $52\, R_{\odot}$ BSG with metallicity $Z=0.01\, Z_{\odot}$, and a $11.9\, M_{\odot}$ and $875\, R_{\odot}$ RSG with $Z=Z_{\odot}$. The mass and radius of stellar progenitors are specified at the end of their evolution. 
    \end{item} 
\end{itemize}

Figure~\ref{density_profiles} shows the radial mass density distribution for the three stellar progenitor profiles that we consider here. The surface layers of both BSG and RSG models predominantly consist of hydrogen and helium, whereas the WR star surface is already stripped off its lighter elements prior to the core-collapse stage. 
The density profile of the WR can be divided into three parts: a power-law profile with index $\alpha \approx -2.5$ up to $R \sim 10^{10}\, {\rm cm}$, followed by a sharp decline in density around $R \sim 3\times10^{10}\, {\rm cm}$ and then a steep power-law tail up to $R \sim 3\times10^{11}\, {\rm cm}$. For the BSG, the density profile is a power-law with index $\alpha \approx -2.5$ up to $R \sim 10^{12}\, {\rm cm}$, followed by a steep drop in density near the edge of the star. In case of the RSG progenitor, the density profile is a power-law with index $\alpha \approx -2.5$ up to $R \sim 2\times10^{11}\, {\rm cm}$ and a shallower profile ($\alpha \approx -2$) up to its radius $R \sim 5\times10^{13}\, {\rm cm}$. 

\section{Analytical model for the jet-cocoon system}
\label{Sec3}
Jet interaction with its surrounding medium plays an important role in determining the propagation velocity of the jet, energy deposited onto the cocoon and therefore the jet breakout criterion. The jet-cocoon interaction decides the morphology of both these components, inside the dense medium and after breaking out of it. Here we discuss a simple analytical model that we adopt to self-consistently determine the properties of the jet-cocoon system.

\subsection{Jet-head propagation} \label{S3.1} 
We consider a magnetized jet with luminosity $L_j$, injected with an opening angle $\theta_j$ into a medium of density $\rho_a(r)$. The jet luminosity is $L_j = \dot{E}_{\rm tot}$ and its isotropic-equivalent one is given by $L_{j,\rm iso} = \dot{E}_{\rm iso} = \dot{E}_{\rm tot}/(\theta_j^2/2)$.
As the jet propagates through the stellar envelope, it forms a bow shock around the jet-head which dissipates jet energy onto the cocoon surrounding the jet. The mildly relativistic cocoon, in turn, exerts pressure on the jet to potentially collimate it and thereby changing its propagation velocity. 
The dynamics of the jet-cocoon system also depends on the outflow magnetization, as the asymptotic jet Lorentz factor is practically limited to $\Gamma_{j,\infty} \lesssim (\sigma_0/\theta_{j,\infty}^2)^{1/3}$ \citep{LB2013}. 
We assume that the entire system is axisymmetric and use the subindices `j', `jh', `c' and `a' to denote quantities related to the jet, the jet-head, the cocoon and the ambient medium, respectively.

\begin{figure*} 
\includegraphics[width=0.32\textwidth]{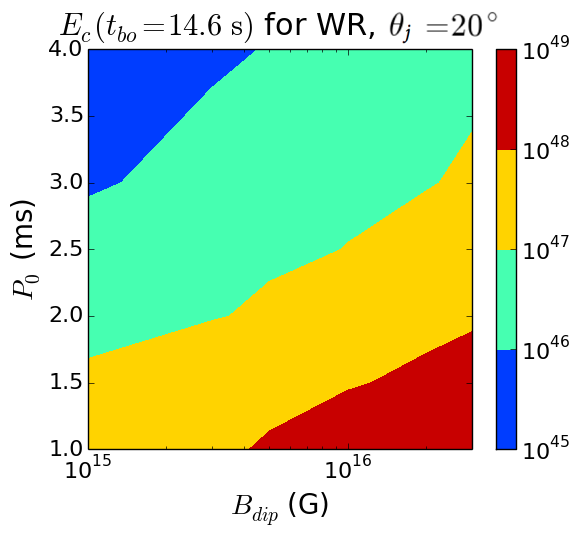}
\includegraphics[width=0.32\textwidth]{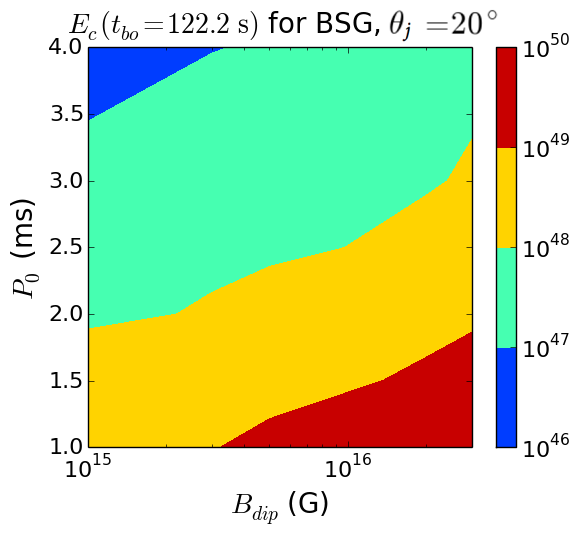}
\includegraphics[width=0.32\textwidth]{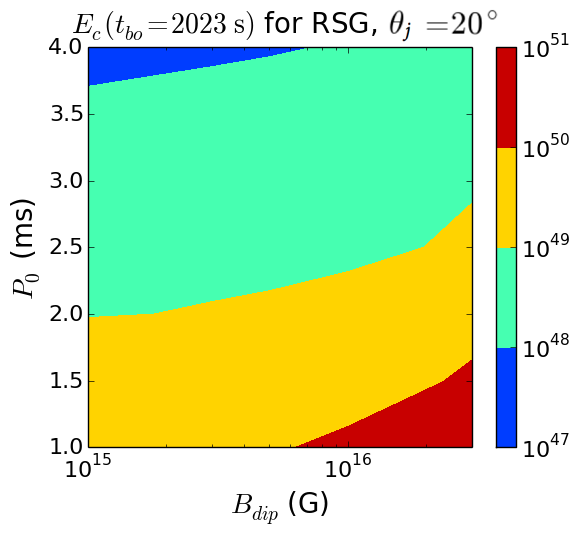}
\vspace{-0.3cm}
\caption{Contours for the energy deposited ($E_c$) by  the magnetized jet onto the cocoon until the breakout time $t_{\rm bo}$ (see Appendix~\ref{App2}), are shown in the $B_{\rm dip}-P_i$ plane. The results are shown here for a fixed jet opening angle $\theta_j=20^{\circ}$ and $\eta=1$. As expected, $E_c(t_{\rm bo})$ is maximised for the central engines with a combination of stronger fields and rapid rotation rates for all three progenitors. 
We find that $E_c(t_{\rm bo})$ is significantly larger for $(15.9M_{\odot},52R_{\odot})$ BSG and $(11.9M_{\odot},875R_{\odot})$ RSG progenitors due to their much longer jet breakout times as compared to $(15.7M_{\odot},5.15R_{\odot})$ WR.
} 
\label{Ec_contours}
\end{figure*}

Figure~\ref{schematic} shows a schematic diagram for:
the jet head, the shocked and unshocked parts of the jet, the inner and outer regions of the cocoon, and the surrounding stellar medium.  
The collimation shock (CS) splits the jet into shocked and unshocked regions, whereas the contact discontinuity separates jet material entering the head from ambient medium and extends to the cocoon to divide it into an inner and outer part. The cocoon expands into the ambient medium behind a shock that extends from the forward shock at the jet head. 
Jets can be either collimated (left panel) or uncollimated (right panel) based on $L_{j,\rm iso}$, $\theta_j$, $\rho_a(r)$ and strength of the CS (see Appendix~\ref{App3} for discussion). Low-power jets, with generally smaller $\sigma_0$, are easier to get collimated inside the star by the ambient medium, as they become slow and cylindrical. In contrast, the bulk of the jet material is unshocked with a conical shape, for magnetized jets with $\sigma_0 \gg 1$. As the jet propagates within the star, it collides with the stellar envelope. A reverse shock then decelerates the jet and a forward shock runs into the stellar envelope, the shocked region is referred to as the jet-head.

The velocity of the jet-head is determined by the ram pressure balance between the shocked jet and the shocked envelope \citep{MW2001,Matzner2003}
\bea
\label{ramP_bal}
h_j \rho_j c^2 \Gamma_{jh}^2 \beta_{jh}^2 + P_j = h_a \rho_a c^2 \Gamma_{h}^2 \beta_{h}^2 + P_a
\eea
where $\Gamma_{jh} = \Gamma_j \Gamma_h (1- \beta_j \beta_h)$ is the relative Lorentz factor between the jet and its head, $\beta_{jh} = (\beta_j -\beta_h)/(1-\beta_j \beta_h)$ is the corresponding relative velocity and $h=1+4P/\rho c^2$ is the specific enthalpy. 
We define the ratio between the jet energy density and rest-mass energy density of the ambient medium as well as the ratio between total jet pressure and energy density of the surrounding medium as
\bea
\Tilde{L} = \frac{L_j}{\pi r_j^2 \rho_a c^3},\
\Tilde{P} = \frac{P_j + B^2/(8\pi \Gamma_j^2)}{\rho_a c^2}
\label{Ltilde}
\eea
where $P_j = L_j/\pi c r_j^2$ is the jet pressure. 

For known $\Tilde{L}$ and $\Tilde{P}$, the general solution for the jet head velocity is given by \citep{LB2013}
\bea
\beta_h = \beta_j \frac{\Tilde{L} - [\Tilde{L}^2 - (\Tilde{L} + \Tilde{P}/\beta_j^2)(\Tilde{L} - \Tilde{P} -1)]^{1/2}}{\Tilde{L} - \Tilde{P} - 1}.
\eea 
For cold ambient matter and strong reverse shock, both $P_j$ and $P_a$ can be ignored. The solution for jet-head velocity then simplifies to $\beta_h = \beta_j/(1 + \Tilde{L}^{-1/2})$ \citep{Matzner2003}. In the limit of a relativistic reverse shock, 
\bea
\Tilde{L}^{1/2} = \frac{1}{\beta_h} - 1 = \left\{
\begin{array}{ll}
2\Gamma_h^2, & \Gamma_j^4 \gg \Tilde{L} \gg 1 \vspace{0.2cm} \\
\beta_h, & \Tilde{L} \ll 1 \\
\end{array}
\right.
\eea
Therefore, the velocity of the jet head depends only on the ambient density near the head and is insensitive to the geometry of the jet below the head.

\subsection{Cocoon properties} \label{S3.2}
Cocoon pressure is sustained by continuous energy inflow from the jet head. At a given time, the total energy deposited in the cocoon is $E_c = \eta L_j(t-r_h/c)$, where $\eta \sim 0-1$ represents the fraction of the energy that flows into the jet head and deposits into the cocoon \citep{Bromberg2011}. For simplicity, we assume $\eta=1$ in our analysis. Assuming that energy density is uniformly distributed within the cocoon, cocoon pressure is determined by the injected energy divided by the cocoon volume (see \citealt{BC1989})
\bea \label{P_c}
P_c = \frac{E_c}{3V_c} = \frac{\eta}{3} \frac{\int L_j (1-\beta_h)dt}{\pi r_c^2 r_h} = \frac{\eta}{3\pi} \frac{\int L_j (1-\beta_h)dt}{(\int \beta_c c\ dt)^2 (\int \beta_h c\ dt)}
\eea
The cocoon geometry can be approximated as a cylinder of height $r_h = c\int \beta_h dt = c\xi_h \beta_h t$ and radius $r_c = c\int \beta_c dt = c\xi_c \beta_c t$. The cocoon's lateral expansion velocity is obtained by balancing $P_c$ with ram pressure of the ambient medium 
\bea
\label{beta_c}
\beta_c = \sqrt{\frac{P_c}{\overline{\rho}_a c^2}}, \ \ \ \ \overline{\rho}_a = \frac{\int \rho_a dV}{V_c} = \xi_a \rho_a
\eea
where $\overline{\rho}_a$ is the mean density of the ambient medium. As the density profiles of stellar progenitors generally follow a power-law dependence $\rho_a \propto r^{-\alpha}$, the coefficients $\xi_a$, $\xi_h$ and $\xi_c$ become constant \citep{MI2013}: $\xi_a = 3/(3-\alpha)$ and $\xi_h = \xi_c = (5-\alpha)/3$. 
For simplicity, $\xi_a$ is obtained assuming a spherical cocoon with radius $r_h$.

Figure~\ref{Ec_contours} shows the contour plots in $B_{\rm dip}-P_i$ plane for total energy deposited by the relativistic head into its surrounding cocoon until the time of jet breakout (see Appendix~\ref{App2} for discussion). 
The results are shown for three progenitors (WR, BSG and RSG), jet-opening angle $\theta_j = 20^{\circ}$ and $\eta=1$. 
For each progenitor, energetic jets originating from PNS with stronger field and rapid rotation deposit more energy into the cocoon material. However, $E_c(t_{\rm bo})$ is not directly affected by the variation in $\theta_j \sim 5-20^\circ$. 
The deposited energy is significantly larger for BSG ($E_c \sim 10^{46}-10^{50}\, {\rm erg}$) and RSG ($E_c \sim 10^{47}-10^{51}\, {\rm erg}$) progenitors in comparison to WR stars ($E_c \sim 10^{45}-10^{49}\, {\rm erg}$), due to their larger $R_*$, and consequently, longer $t_{\rm bo}$.

\begin{figure*} 
\includegraphics[width=0.32\textwidth]{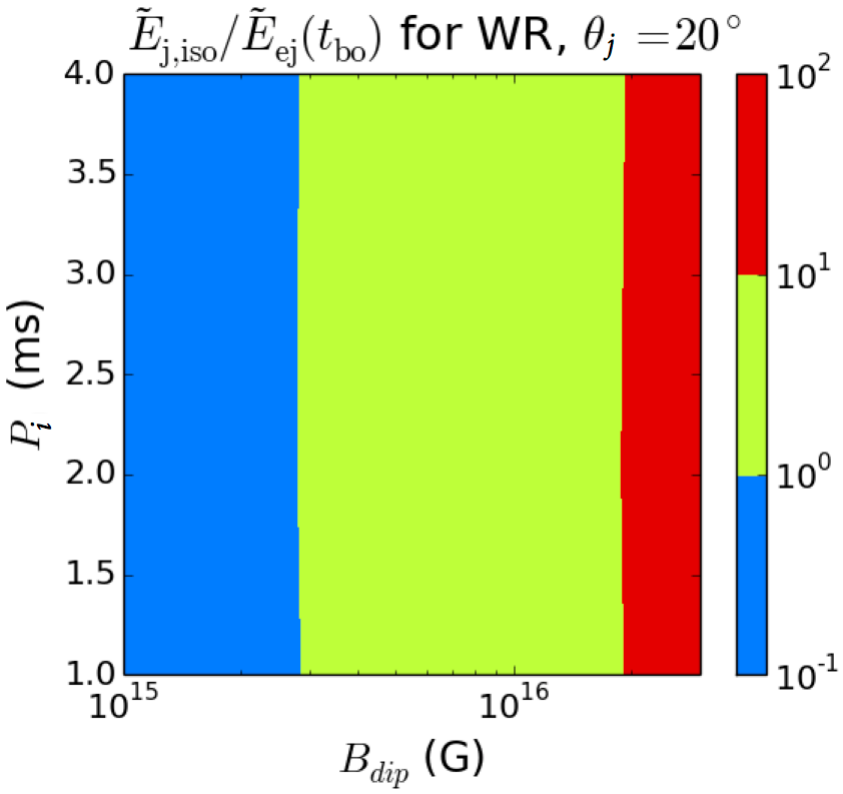} 
\includegraphics[width=0.32\textwidth]{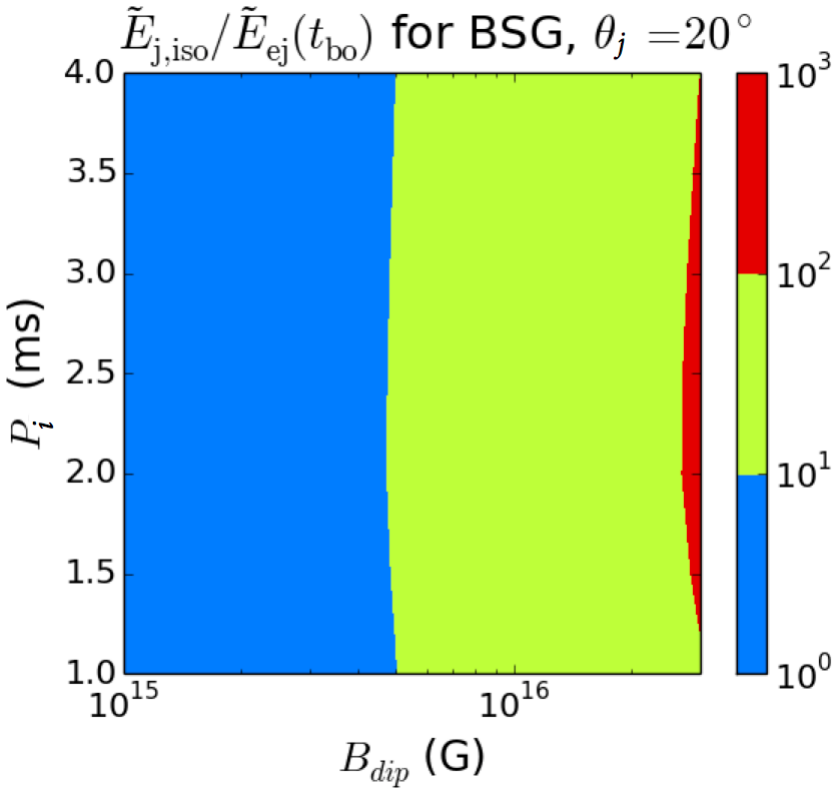} 
\includegraphics[width=0.32\textwidth]{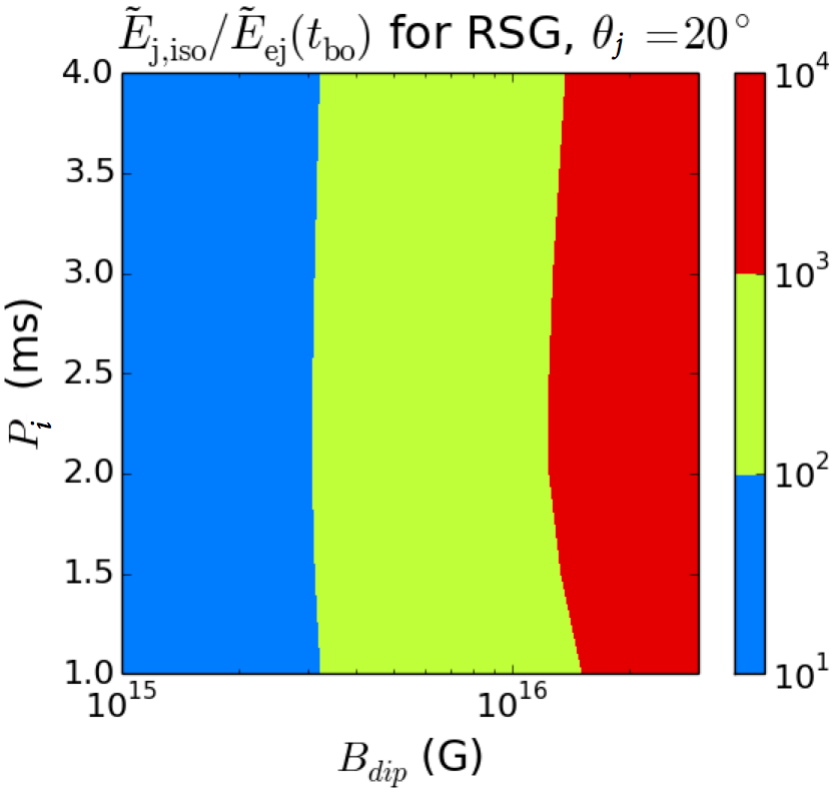}
\vspace{-0.3cm}
\caption{Contours for the ratio between jet isotropic-equivalent energy and effective energy deposited into the cocoon, both computed at the jet breakout time, are shown in $B_{\rm dip}-P_i$ plane. The results are shown here for the WR [left panel], BSG [center panel] and RSG [right panel] progenitor, respectively. We find that $\tilde{E}_{\rm j,iso}/\tilde{E}_{\rm ej}(t_{\rm bo})$ has a relatively weak dependence on $P_i$ for all three progenitors. While jets from compact objects with $10^{15} \lesssim B_{\rm dip}/{\rm G} \lesssim 3\times10^{16}$ and $1 \lesssim P_i/{\rm ms} \lesssim 4$ can successfully break out ($\tilde{E}_{\rm j,iso}/\tilde{E}_{\rm ej} \gtrsim 1$) from the less dense BSG/RSG progenitors, outflows arising from PNS with relatively weaker fields $B_{\rm dip} \lesssim 3\times10^{15}\, {\rm G}$ get choked ($\tilde{E}_{\rm j,iso}/\tilde{E}_{\rm ej} < 1$) inside the envelope of WR stars.
} 
\label{Ej_Ec_comp}
\end{figure*}

\subsection{Collimation of the jet} \label{S3.3}
Jet collimation occurs due to the formation of an oblique shock at the jet base, which deflects the jet flow-lines and generates a pressure that counterbalances $P_c$. To maintain the required pressure, the shock curves towards the jet axis until it converges at some height, above which collimation is complete. 
While $\tilde{L}$ determines the evolution of the jet-cocoon system, $\tilde{L} \theta_j^{4/3}$ distinguishes between collimated and uncollimated jets \citep{Bromberg2011}. $\tilde{L}$ can vary with jet propagation, even if $L_j$ remains unchanged, depending on the density profile and jet's cross section. 
The shape of the initially conical jet changes to cylindrical after collimation and the CS sets jet head's cross section to be much smaller than the cross section of uncollimated jet. 
Consequently, the jet applies a larger ram pressure on the head to push it to higher velocities further reducing the energy flow rate into the cocoon.  
The cocoon's height then increases at a faster rate, leading to a larger volume and decrease in $P_c$. Therefore, there is a upper limit to $\beta_h$ above which $P_c$ becomes insufficient to effectively collimate the jet.

The CS geometry is determined by the pressure balance between the jet and the cocoon \citep{Bromberg2011} 
\bea
h_0 \rho_0 c^2 \Gamma_0^2 \beta_0^2 {\rm sin}^2 \psi + P_i = P_c
\eea
where the first term is the jet ram pressure normal to the shock surface and the subscript `0' stands for the unshocked jet. 
Here $\psi$ denotes the angle between the direction of the relativistic outflow and the CS surface.
As the jet internal pressure falls off as $P_i \propto r^{-4}$ with increase in its size, the $P_i$ term can be neglected. For small incident angle and to the first order,
\bea  
{\rm sin}\psi = \frac{R_{s}}{r} - \frac{dR_{s}}{dr} = r\frac{d}{dr}\left(\frac{R_{s}}{r}\right)
\eea
where $R_{s}$ is the cylindrical radius of the shock position. 
Assuming that $\beta_0 \approx 1$ and $L_j \approx h_0 \rho_0 c^3 \Gamma_0^2 (\pi r^2 \theta_j^2)$, one can integrate to obtain the CS geometry, $z_{\rm cs} = \theta_j r (1+ Az_*) - \theta_j A r^2$, where $A = \sqrt{\pi c P_c/L_j \beta_0}$ is evaluated using constant $P_c$.  
The CS expands to maximum size at $(dz_{\rm cs}/dr)|_{r=z_{max}}=0$, and converges to $r_s(r=z_{\rm cs})=0$ where the maximally expanding position $z_{max}$ and the converging position $z_{\rm cs}$ are  
\bea \label{zhat}
z_{\rm cs} = 2 z_{\rm max} = (1/A) + z_* \sim 1/A = \sqrt{\frac{L_j \beta_0}{\pi c P_c}}
\eea
Here we assume that the CS is initially small with $z_* \ll 1/A$ (see e.g., \citealt{MI2013}). 
At a given time post core-collapse, both $r_h$ and $z_{\rm cs}$ tend to be larger for jets that originate from PNS with stronger field and rapid rotation. 
For outflows with similar $(B_{\rm dip},P_i)$, less dense external medium (e.g., BSG and RSG progenitors) leads to a larger $r_h$ and the CS also converges at larger distance from the central engine.

\section{Jet choking and stability}
\label{Sec4}
Observed jet duration is the difference between engine’s operation time ($t_{\rm eng}$) and jet's breakout time ($t_{\rm bo})$. For a successful jet breakout, the central engine has to be active for at least the threshold activity time, $t_{\rm th} = t_{\rm bo} - R_*/c$, where $R_*/c$ is the light crossing time of the star. Here we discuss the criteria for successful jet breakout and also the stability of magnetized jets relative to current-driven instabilities.

\subsection{Jet choking criteria} \label{S4.2}
Below a critical jet isotropic luminosity $L_{j,\rm iso} = L_j/(\theta_j^2/2)$, both hydrodynamic and Poynting flux dominated outflows may fail to produce stable jets, which are instead choked inside the star (see e.g., \citealt{Metzger2011a}). For such choked jets, the cocoon is the only component that breaks out from the star and spreads quasi-spherically forming an entirely different structure from jets that typically break out. 

The central engine needs to generate a minimum amount of energy to push the jet out of the star. This corresponds to minimal engine activity time, $t_{\rm eng} \gtrsim t_{\rm th} = t_{\rm bo} - R_*/c$, necessary for a successful jet breakout. 
If the jet head is non-relativistic during the entire crossing, $t_{\rm bo} \gg R_*/c$ and $t_{\rm th} \approx t_{\rm bo}$. For a relativistic jet head, the threshold time is $t_{\rm th} = (1.1\, {\rm sec}) L_{j,49}^{-1/3}M_{*,15M_{\odot}}^{1/3}R_{L,7}^{2/3}$. 
The jet can get choked within the stellar envelope due to two possibilities: (a) the central engine stops at $t < t_{\rm th}$ before the jet manages to exit the star, (b) the isotropic jet power is less than the minimum requirement.  
Recently, \citet{GN21} derived a breakout criterion that just depends on the jet opening angle and the jet to ejecta energy ratio. In particular, for strongly magnetized jets where the engine activity time $t_{\rm eng}$ typically exceeds the delay time for jet launch since core-collapse, $\tilde{E}_{\rm j,iso} \gtrsim \tilde{E}_{\rm ej} \approx 40 E_{\rm ej,tot} \theta_j^2$, evaluated at $t_{\rm bo}$. Here, $\tilde{E}_{\rm j,iso} = \int_0^{t_{\rm bo}}L_{j,\rm iso}dt$ and $E_{\rm ej,tot} \approx E_c(t_{\rm bo})$.  

Figure \ref{Ej_Ec_comp} shows contour plots in $B_{\rm dip}-P_i$ plane for $\tilde{E}_{\rm j,iso}/\tilde{E}_{\rm ej}$ evaluated at $t_{\rm bo}$. Results are shown for three progenitors (WR, BSG and RSG) and for $\theta_j=20^{\circ}$. 
As $L_{j,\rm iso} \propto \theta_j^{-2}$ for the time-independent $L_j$ at later times (see Figure~\ref{sigma0_Edot_t}), $\tilde{E}_{\rm j,iso}/\tilde{E}_{\rm ej} \propto \theta_j^{-4}$ increases sharply for smaller $\theta_j \sim 5-10^{\circ}$ aiding jet breakout, and we choose $\theta_j=20^{\circ}$ to provide a conservative estimate. 
Irrespective of the density profile considered, we find that $\tilde{E}_{\rm j,iso}/\tilde{E}_{\rm ej}(t_{\rm bo})$ has a weak dependence on $P_i$, which therefore does not affect the breakout condition. Magnetized outflows arising from PNS with weaker fields $B_{\rm dip} \lesssim 3\times10^{15}\, {\rm G}$ can get choked inside the envelope of WR stars as $\tilde{E}_{\rm j,iso}/\tilde{E}_{\rm ej}(t_{\rm bo}) \lesssim 1$. In contrast, for comparable ranges of $B_{\rm dip}$ and $P_i$, jets tend to successfully break out from the less dense BSG and RSG progenitors.

\subsection{Stability of the magnetized jet} \label{S4.3}
Prior to jet breakout, instabilities may develop along the jet-cocoon boundary from the collimation point up to the jet head and lead to a diffused structure that separates the jet from the cocoon, termed as the jet-cocoon interface (JCI, \citealt{Gottlieb2020a}). The growth of instabilities incites efficient mixing of the jet and cocoon material along JCI. Magnetic fields inhibit the growth of local hydrodynamic instabilities (see e.g., \citealt{MM2019}).  
magnetization required to stabilise the jet depends on various jet parameters such as the initial opening angle and jet power. Wider and/or low-power jets require stronger fields for stabilisation whereas narrower and/or high-power jets are generally more stable and therefore the required $\sigma_0$ is smaller. 
Even a sub-dominant magnetization ($0.01 < \sigma_0 < 0.1$) can stabilise the jet against hydrodynamic instabilities on the jet-cocoon boundary, allowing the jet to maintain a larger fraction of its original energy \citep{Gottlieb2021a}.  
However, in highly magnetized jets ($\sigma_0 > 1$), current-driven instabilities such as kink may emerge and potentially render the jet structure globally unstable \citep{Matsumoto2021}. Narrow jets are most susceptible to kink instability which excites large-scale helical motions that can strongly distort the jet and thus trigger violent magnetic dissipation \citep{Eichler1993,Lyubarskii1999,GS2006}. Poynting-flux dominated jets can survive the crossing of the star if the crossing time is shorter than the growth time of the instability \citep{Bromberg2014}.

\begin{figure}
\centering
\includegraphics[width=0.92\columnwidth]{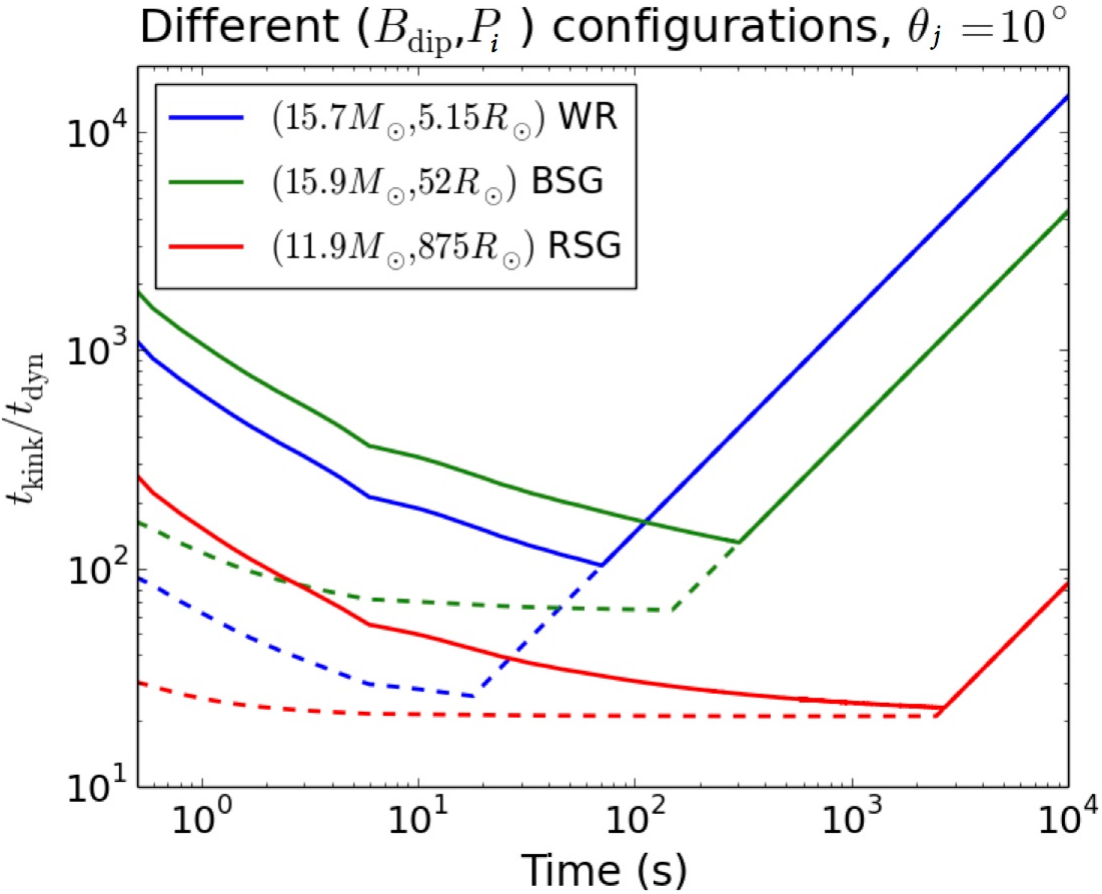}
\vspace{-0.3cm}
\caption{Comparison between the growth timescale for kink instability and system dynamical timescale for jets with $(B_{\rm dip},P_i)=(10^{15}\, {\rm G},2\, {\rm ms})$ [solid curves] and $(10^{16}\, {\rm G},1\, {\rm ms})$ [dashed curves]. The blue, green and red curves are shown for WR, BSG and RSG progenitor, respectively. Jets originating from protomagnetars with stronger fields and rapid rotation rates tend to be more stable at earlier times. 
} 
\label{t_kink}
\end{figure}

While outflows with purely toroidal fields are most sensitive to disruption by kink instability \citep{Mignone2010,ON2012}, toroidal fields comparable to the poloidal field can result in a more stable flow \citep{Lyubarsky2009}. 
The full growth of kink instability requires a timescale which is comparable to a few light crossing times of the jet width (see \citealt{LB2013}). The total time necessary for the full development of the instability is then $t_{\rm kink}^{\prime} \sim 10fr_j/c$, where $f \sim 0.5-1$ is a numerical factor \citep{Mizuno2009,Mizuno2012}. As the unstable perturbation propagates with the jet plasma in the observer frame, the instability disruption timescale is
\bea \label{tkink}
t_{\rm kink} = \Gamma_j t_{\rm kink}^{\prime} \sim 100f \frac{r_j^2}{cR_L},
\eea
where $R_L=c/\Omega$ is the light cylinder radius. 
This time should be compared with the dynamical time $t_{\rm dyn} = {\rm min}(r_h/c,R_*/c)$ available for the instability to grow in the jet.

Figure~\ref{t_kink} shows a comparison between the time required for kink instability to grow in the jet and the expansion timescale of the magnetized jet for two PNS configurations.  
Results are shown for the WR, BSG and RSG progenitors, and for $\theta_j = 10^{\circ}$. The gradual increase in $t_{\rm kink}/t_{\rm dyn}$ at later times occurs due to the rapid jet expansion post breakout ($r_h \gtrsim R_*$). Magnetized jets arising from PNS with stronger fields and rapid rotation rates are generally more stable to the kink instability. While $t_{\rm kink}/t_{\rm dyn}$ is somewhat comparable for all three progenitors considered, jets with wider opening angles tend to be more stable as $t_{\rm kink} \propto r_j^2$. 
\citet{Bromberg2014} performed a similar analysis to test the stability of the magnetized jet under the assumption that it is non-rotating and moves upward rigidly. They showed that the jet is expected to be stable as its typical width is $\sim {\rm few}\ 10{\rm s}$ of $R_L$, so the kink instability does not have enough time to develop in the jet before it breaks out of the star.

\section{High-energy neutrino emission} \label{Sec5}
We consider magnetized jets with time-varying luminosity that arise from PNS central engines, to calculate neutrino spectra during the phase when the jet propagates within a stellar progenitor. This provides a better estimate of high-energy neutrino production that depends on the dissipation radius, which in turn is a function of the central engine parameters. Neutrinos are produced from pion and muon decay which originate from $p\gamma$ interactions. We assume that the protons are accelerated around the termination shock, where for highly magnetized outflows, strong magnetic dissipation is expected to occur as in pulsar wind nebulae. For the target photons, we consider photons that originate from the jet head and leak into the jet.

The central engine parameters that we focus on are: $B_{\rm dip} \sim 10^{15}-10^{16}\, {\rm G}$, $P_i \sim 1-2\, {\rm ms}$ and $\theta_j=10^{\circ}$.
The typical breakout time and jet Lorentz factors $\Gamma_j$ for the chosen parameters are summarized in Table \ref{NuProgenitorParameters}. 
For these parameters, the jet remains uncollimated until breakout and neutrino production originates from the interactions of high-energy particles. 

Shocks can be collisionless in the presence of large magnetic fields but the dominant fraction of the outflow energy is magnetic, so we consider magnetic reconnection as the plausible energy dissipation mechanism \citep{Usov1992,Drenkhahn2002}. 
While some half of the dissipated energy gets converted into jet kinetic energy, the remaining half is utilised to accelerate the particles. Once the outflow magnetization reduces to $\sigma_0 \sim 1$ and the jet stops accelerating, magnetic reconnection no longer remains efficient. While particles can be accelerated up to the magnetic dissipation radius $R_{\rm mag}$, where the outflow Lorentz factor finally saturates, we stop their injection at the stellar radius.

Assuming that the magnetic field is predominantly toroidal far from the PNS surface and shock propagates perpendicular to the field, \citet{KC1984} showed that the Rankine-Hugonoit relations can be simplified to obtain the particle temperature in the jet-comoving frame 
\bea
kT^{\prime} = m_e c^2\left\{
\begin{array}{ll}
\frac{1}{\sqrt{18}}(1-2\sigma_0), & \sigma_0 < 0.395 \vspace{0.2cm} \\
\frac{1}{8\sqrt{\sigma_0}}\left(1-\frac{0.297}{\sigma_0}\right), & \sigma_0 \gtrsim 0.395 \\
\end{array}
\label{JumpConditions}
\right.
\eea
While large $\sigma_0$ shocks are generally weak, strong shocks effectively approach the classical hydrodynamical limit for small $\sigma_0$. For the range of PNS parameters considered here, we obtain $T^{\prime} \sim 10^{7}-10^{9}\, {\rm K}$. In the rest frame of the shocked jet-head, the photon density and energy of individual photons are boosted by a factor of $\Gamma_{jh} = \Gamma_j\Gamma_h (1-\beta_j\beta_h) \sim 10^3-10^5$, which implies that nuclei are completely dissociated into protons within the outflow. 
These protons are accelerated and we assume a spectrum (number of protons per unit comoving proton energy $\varepsilon_p^\prime$), $dN_p/d\varepsilon'_p \propto \varepsilon_p^{\prime -s}$, where primes indicate quantities in the jet-comoving frame and $s$ is the power-law spectral index. The jet becomes highly magnetized over time, for which hard spectra with $1 \leq s \leq 2$ are motivated by numerical simulations (see e.g., \citealt{Guo_2016,Werner:2014spa}). While $s\approx2$ for $\sigma_0=10$, the proton spectrum becomes harder as $\sigma_0$ increases: $s\approx 1.5$ for $\sigma_0=50$ and $s\approx 1$ for highly magnetized outflows with $\sigma_0\gg 1$ (see e.g., \citealt{Kagan:2014hea}). For this reason, we will consider the cases with $s=1$ and $s=2$. 

The magnetic field strength in the outflow is given by $B^\prime = \sqrt{2\epsilon_B L_{j,\rm iso}/r_h^2 \Gamma_h^2 c}$, where $\epsilon_B=0.3$ is the fraction of the isotropic luminosity converted to magnetic field energy.
\citet{Petro2019} performed large-scale 2D particle-in-cell (PIC) simulations to determine the mean electron energy in the post-reconnection region, for a range of outflow magnetization and particle temperatures. Other PIC simulations for electron and ion energy spectra, including 3D, have also been explored in the literature \citep{Ball_2018,Zhang:2021akj}. 
Here we adopt their model to estimate the mean energy of the electrons from
\begin{equation}
\langle\gamma_{e}-1\rangle = \sqrt{\sigma_0}\left(1 + \frac{4kT'_e}{m_e c^2}\right)\left(1 + \frac{\sigma_{e,h}}{30}\right),
\end{equation}
where $T'_e$ is the electron temperature and $\sigma_{e,h}$ is the pair plasma magnetization. 
It should be noted that the precise value of pair plasma magnetization depends on the energies of particles as well as the pair multiplicity (see \citealt{Petro2019}, for more exact estimates).

The maximum proton energy is determined by the balance between acceleration time $t^\prime_{p,\text{acc}} = \eta_{\rm acc}\varepsilon'_p/(eB'c)$ and cooling time $t^\prime_\text{cool}=(t_{\rm ad}^{\prime -1}+t_{\rm syn}^{\prime -1}+t_{p\gamma}^{\prime -1})^{-1}$, where $t_{\rm ad}^\prime = r_h/\Gamma_{jh}c$ is the adiabatic timescale and $\eta_{\rm acc}\sim1$ is assumed. 
We multiply the proton spectrum by the suppression factor $\exp(-\varepsilon_p^\prime/\varepsilon_{p,{\rm max}}^\prime)$, to account for the maximum proton energy. 
The minimum proton energy is $\varepsilon_{p, {\rm min}}^\prime = \sigma_0^{1/2} m_p c^2$ and we normalize the injection spectrum such that the energy injection rate is equal to the jet isotropic-equivalent kinetic luminosity $\epsilon_p L_{j,{\rm iso}}$, where $\epsilon_p=0.3$ is the energy fraction given to cosmic rays. We choose $\epsilon_p=0.3$ based on our assumption of equipartition of the total energy such that $\epsilon_p \sim \epsilon_e \sim \epsilon_B$.

\begin{table}
\centering
\caption{Characteristic range of parameters for WR, BSG and RSG progenitors. $\Gamma_j$ is evaluated at $t_{\rm bo}$.}
\def\arraystretch{1.3}
\begin{tabular}{|c|c|c|}
\hlinewd{0.75pt} 
Progenitor &  $t_{\rm bo} (s)$ & $\Gamma_j (t_{\rm bo})$ \\ \hlinewd{0.25pt} \hline
WR  &  $10 - 60$ & $20-60$ \\
BSG & $100 - 200$ & $10^3-10^4$\\
RSG & 2000 & $10^5$\\
\hlinewd{0.25pt} \hline
\end{tabular}
\label{NuProgenitorParameters}
\end{table}

\begin{figure*}
\includegraphics[width=0.50\textwidth]{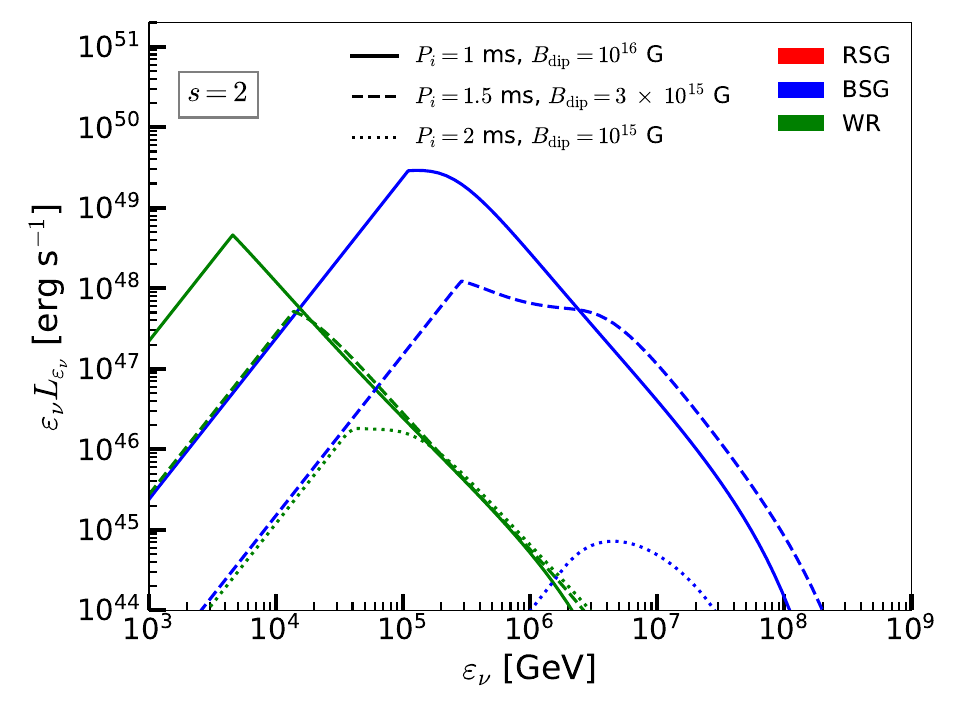}
\includegraphics[width=0.49\textwidth]{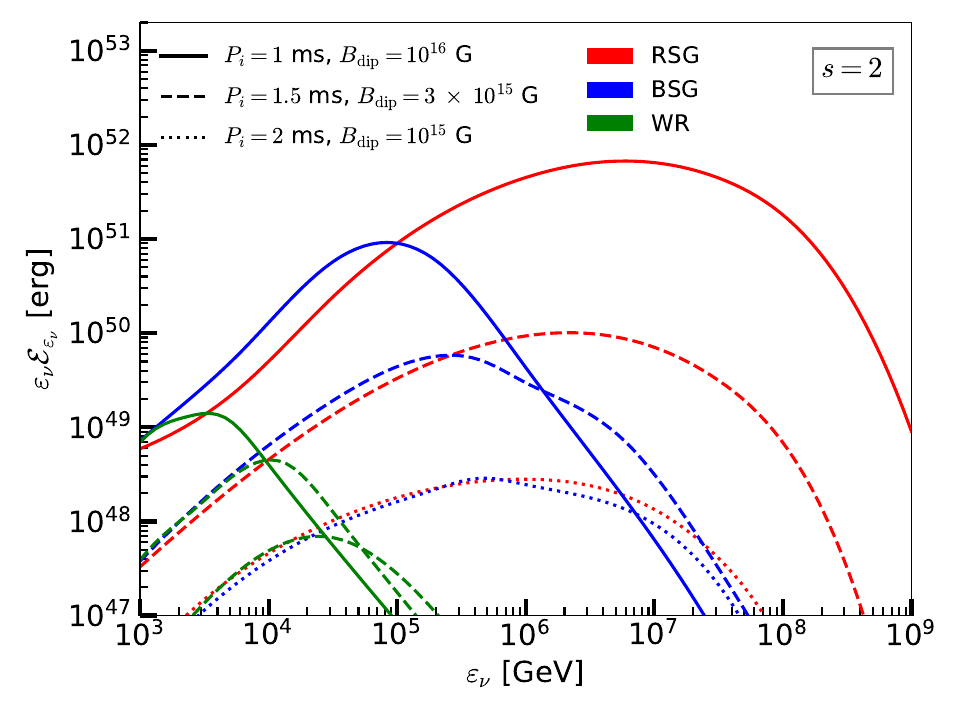}
\includegraphics[width=0.50\textwidth]{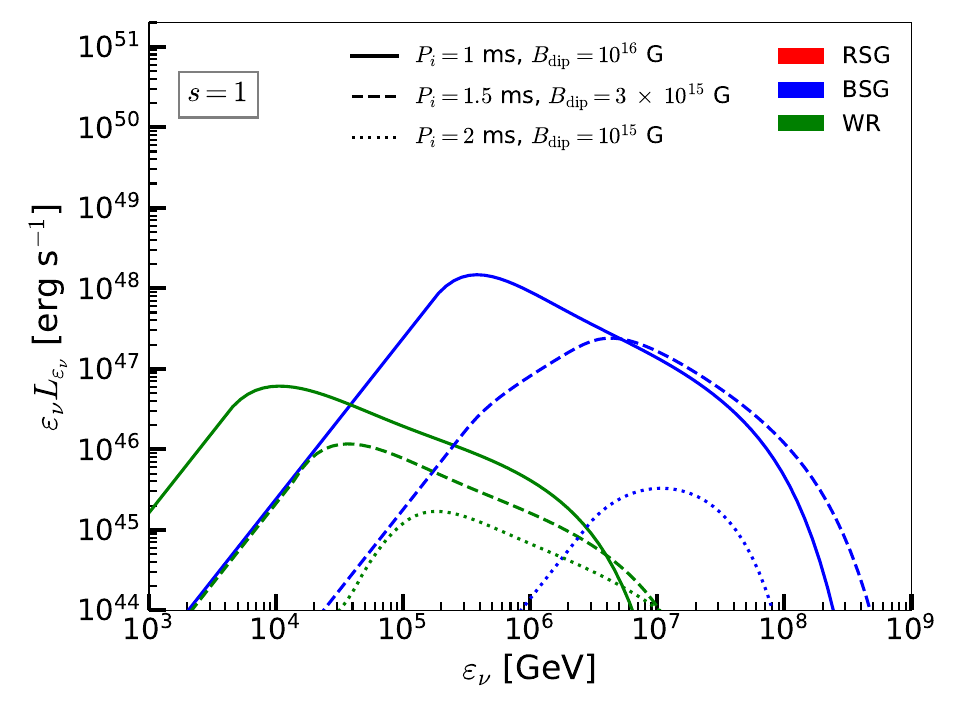}
\includegraphics[width=0.49\textwidth]{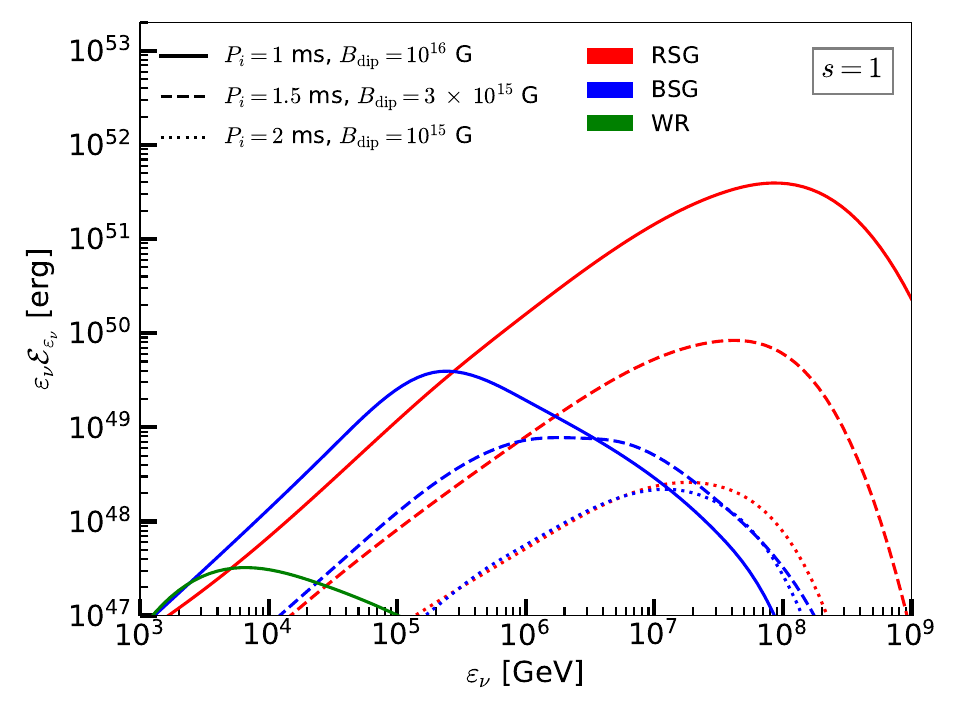}
\vspace{-0.3cm}
\caption{\textbf{Left panels:} Neutrino spectra in the engine frame at the breakout time, summed over all three flavors are shown for magnetized outflows with $(B_{\rm dip},P_i)=(10^{16}\, {\rm G},1\, {\rm ms})$ [solid curves], $(3\times10^{15}\, {\rm G},1.5\, {\rm ms})$ [dashed curves] and $(10^{15}\, {\rm G},2\, {\rm ms})$ [dotted curves]. The green, blue and red curves are shown for the $(15.7M_{\odot},5.15R_{\odot})$ WR, $(15.9M_{\odot},52R_{\odot})$ BSG and $(11.9M_{\odot},875R_{\odot})$ RSG progenitor, respectively. The spectra for RSG are not shown because at the breakout time the photons are non-thermal, causing a rapid decline in $f_{p\gamma}$ and the resulting spectra falls below $10^{44}$ erg s$^{-1}$. The top and bottom panels correspond to an $\varepsilon_p^{\prime -s}$ proton injection spectrum with $s=2$ and $s=1$, respectively.
\textbf{Right panels:} Neutrino fluence i.e. the neutrino spectrum integrated up to $t=t_{\rm bo}$ is shown for the same $(B_{\rm dip},P_i)$ configurations and progenitor models. 
}
\label{NeutrinoSpectra}
\end{figure*}

The protons will interact with the ambient photons to create pions, with timescale $t_{p\gamma}$ given by the formula
\begin{equation}
t_{p\gamma}^{-1}(\varepsilon_p^\prime)=\frac{c}{2\gamma_p^2}\int_0^\infty \frac{d\varepsilon_\gamma^\prime}{\varepsilon_\gamma^{\prime 2}} \frac{dn_\gamma^\prime}{d\varepsilon_\gamma^\prime}\int_0^{2\gamma_p\varepsilon_\gamma^\prime}d\bar{\varepsilon}_\gamma \bar{\varepsilon}_\gamma \sigma_{p\gamma}(\bar{\varepsilon}_\gamma)\kappa_{p\gamma}
\label{PhotopionProductionFormula}
\end{equation}
where $\sigma_{p \gamma}$ is the photomeson production cross section obtained from \citet{Murase:2005hy}, $\kappa_{p\gamma}$ is the proton's inelasticity, $\gamma_p=\varepsilon_p^\prime/m_p$ is the proton's Lorentz factor, $dn_\gamma/d\epsilon'_\gamma$ is the comoving target photon density per energy and $\bar{\varepsilon}_\gamma$ is the photon energy in the proton rest frame.
The photon spectrum is determined as follows. The Thomson optical depth for the jet head is $\tau_{\rm h} \approx \theta_j L_{j,\rm iso} \sigma_T /4\pi r_h \Gamma_h\Gamma_j m_p c^3$, where $\sigma_T$ is the Thomson cross section. The comoving jet head thickness is $\sim \theta_j r_h$ \citep{MW2001}. When $\tau_{\rm h}\gg 1$, photons at the termination shock are thermalized to the temperature $T_\gamma^\prime$ given by $aT_\gamma^{\prime 4} = B^{\prime 2}/8\pi$, and a fraction $1/\tau_{\rm h}$ of these photons leaks into the jet for subsequent $p\gamma$ interactions. Note that this is the temperature in the far-downstream region, which is different from the temperature $T^\prime$ in the immediate downstream, given by the hydrodynamical jump conditions in equation \eqref{JumpConditions}. In general, we can have $T^\prime >T_\gamma^\prime$.

When $\tau_{\rm h}<1$, we use a non-thermal photon spectrum that follows a broken power law \citep{Murase:2006mm}. In the termination shock frame, we have $dn_\gamma^\prime /d\varepsilon_\gamma^\prime = n_b (\varepsilon_\gamma^\prime/\varepsilon_b)^{-1.5}$ for $\varepsilon_{\rm min} < \varepsilon_\gamma^\prime \leq \varepsilon_b$ and $n_b(\varepsilon_\gamma^\prime/\varepsilon_b)^{-2.2}$ for $\varepsilon_b\leq \varepsilon_\gamma^\prime \leq \varepsilon_{\rm max}$. When $\varepsilon_b>\varepsilon_{\rm max}$, we use $dn_\gamma^\prime /d\varepsilon_\gamma^\prime = n_b (\varepsilon_\gamma^\prime/\varepsilon_b)^{-1.5}$ for $\varepsilon_{\rm min}\leq \varepsilon_\gamma^\prime \leq \varepsilon_{\rm max}$. The photon break energy $\varepsilon_b$ is calculated as $\varepsilon_b = \langle \gamma_e\rangle^2\Gamma_j e B^\prime / m_e c$. We also choose $\varepsilon_{\rm min}=1~{\rm eV}$, while the maximum photon energy is limited by $\varepsilon_{\rm max} \approx  0.15~{\rm TeV}~ \Gamma_{j,3}$, with $\Gamma_{j,3} = \Gamma_j/10^3$ \citep[e.g.,][]{Kashi2016}.
The normalization $n_b$ is such that
\begin{equation}
\int_{\varepsilon_{\rm min}}^{\varepsilon_{\rm max}} d\varepsilon_\gamma^\prime\varepsilon_\gamma^\prime \frac{dn^\prime_\gamma}{d\varepsilon_\gamma^\prime} = \frac{\epsilon_B L_{j,\rm iso}}{4\pi r_h^2\Gamma_h^2 c}.
\end{equation}

The fraction of photons that enter the unshocked jet and contribute to photopion production is estimated by $f_{\rm esc} = {\rm min}(1,\tau_{\rm h}^{-1})$. The photon density entering Eq. \eqref{PhotopionProductionFormula} is thus the density at the termination shock (for its corresponding $\tau_h$) multiplied by $f_{\rm esc}$ and boosted to the jet frame by the factor $\Gamma_{jh}$.
The calculation of $t_{p\gamma}^\prime$ is then used to define the effective optical depth
\begin{equation}
f_{p\gamma}= {\rm min}(1, t'_{\rm cool} {t'}_{p\gamma}^{-1}).
\label{fpgamma}
\end{equation}

Pions and muons lose their energies with their associated cooling rates $t_{\rm cool}^{\prime -1} = t_{\rm ad}^{\prime -1}+t_{\rm syn}^{\prime -1}+t_{\rm IC}^{-1}$, where $t_{\rm syn}^{\prime}=6\pi m^4c^3/\sigma_Tm_e^2Z^4\varepsilon^\prime B'^2$ is the synchroton cooling timescale in the comoving frame for a particle of mass $m$, comoving energy $\varepsilon^\prime$ and charge $Ze$ and $t^\prime_{\rm IC}= 3m^2/4c\sigma_{\rm IC}\Gamma_{jh}^2U_{\rm rad}\varepsilon'$ is the Inverse Compton (IC) timescale, where $U_{\rm rad}$ is the radiation energy density and $\sigma_{\rm IC}$ is the IC cross section, which also accounts for the Klein-Nishina suppression at the highest energies. In the presence of cooling, the pion and muon fluxes are each modified by the suppression factor $f_{\rm sup}=1-\exp(-t'_\text{cool}/t'_\text{dec})$, where $t'_\text{dec}$ is the decay timescale in the jet comoving frame. 

Neutrinos are mainly produced via $p\gamma$ interactions inside the unshocked jet where the baryon density is so small that the jet is not radiation-dominated. 
We get the per-flavor neutrino spectrum in jet frame from \citep{Kimura:2022zyg}
\begin{equation}
{\varepsilon'_\nu}^{2}\frac{dN'_\nu}{d\varepsilon'_\nu}\approx \frac{1}{8}f_\text{sup}f_{p\gamma}{\varepsilon'_p}^2\frac{dN'_p}{d\varepsilon'_p},
\label{NuSpectrumFormula}
\end{equation}
where $f_{\rm sup} = f_{\rm sup}^\pi$ for $\nu_\mu$ produced during pion decay and $f_\text{sup} = f_\text{sup}^\pi f_\text{sup}^\mu$ for neutrinos originating from muon decay. The spectrum is then boosted to the observer frame and injected at the corresponding radius, giving $dN_\nu/d\varepsilon_\nu$, where $\varepsilon_\nu$ is the neutrino energy in the engine frame. 
The spectrum $dN_\nu/d\varepsilon_\nu$ becomes flat for energies below the corresponding minimum pion energy. This happens if $\varepsilon_\nu^\prime<\varepsilon^\prime_{p, {\rm min}}/20$, or when $\varepsilon^\prime_p$ is low enough to significantly suppress $t_{p\gamma}^{-1}$ as a result of the energy threshold of $\sigma_{p\gamma}$. Finally, we propagate the spectrum through vacuum to get $dN_\nu/dE_\nu$.

Neutrino propagation is simulated using {\tt nuSQuIDS} \citep{Arguelles21}, while neutrino oscillation parameters are fixed to the best fit values provided by {\tt NuFIT 2021} \citep{Esteban2020}. The propagation is discontinued once the neutrinos reach $R_*$. For a source distance $D$, the neutrino flux is given by $\phi_\nu = (1/4\pi D^2)dN_\nu/dE_\nu$, while the neutrino fluence $\Phi_\nu$ is the time integral of $\phi_\nu$ until $t_{\rm bo}$. We adopt a source distance of $D=100\, {\rm Mpc}$ for our calculations. For the parameters that we have chosen, $\tau_h>1$ throughout the majority of the neutrino emission phase, so the photopion production is caused mostly by a thermal photon spectrum. In the case of RSG progenitors, $\Gamma_j\sim 10^4 - 10^5$ after $\approx 400$ s, which in turn reduces $\tau_h$ significantly during late emission. The non-thermal photons after this time have maximum energies $\sim 10$ TeV and their number density $n_\gamma$ drops significantly when compared to the number density of thermal photons. This decline results in  $f_{p\gamma}\ll 1$ and very few neutrinos are produced.

The neutrino spectrum computed at the breakout time is shown in the left panels of Figure~\ref{NeutrinoSpectra}, for different $(B_{\rm dip},P_i)$ configurations and progenitor models, assuming both $\varepsilon_p^{\prime -2}$ and $\varepsilon_p^{\prime -1}$ proton spectra. We first look at $\varepsilon_p^{\prime -2}$ injection spectra (top panels) and then compare them against their $\varepsilon_p^{\prime -1}$ counterparts (bottom panels). 
We defined $L_{\varepsilon_\nu} = \varepsilon_\nu dN_\nu/d\varepsilon_\nu$ and $\mathcal{E}_{\varepsilon_\nu} = \int_0^{t_{\rm bo}} dt L_{\varepsilon_\nu}$ as the time integral of $L_{\varepsilon_\nu}$.
As the flux is evaluated at the time of jet breakout, it is equal to the injected neutrino flux and no neutrino attenuation is present. In the case of RSG progenitors, the non-thermal photon spectrum at $t_{\rm bo}$ produces a neutrino spectrum with $\varepsilon_\nu L_{\varepsilon_\nu}< 10^{44}$ erg s$^{-1}$ and is therefore not shown in the panels. For the other progenitors, there are three prominent features: spectral break due to $\varepsilon_{p,{\rm min}}^\prime$, pion and muon cooling, each with a suppression factor proportional to $\varepsilon_\nu^{-2}$ for $t_{\rm cool}^\prime\ll t_{\rm dec}^\prime$, and proton spectrum suppression when $t_{\rm cool}^\prime \ll t_{p,{\rm acc}^\prime}$. 
We find that stronger fields and rapid rotation rates will lead to larger neutrino fluxes, which is related to the dependence of $L_{j,\rm iso}$ on these parameters. 
In the case of WR progenitors, $t_{\rm bo}\lesssim 100\, {\rm s}$ and $\Gamma_j \lesssim 10^4$, so we do not see the effects of $\varepsilon_{p,{\rm min}}^\prime$ for $\varepsilon_\nu \gtrsim 1\, {\rm TeV}$. For BSGs, however, the break in the spectrum is caused by $f_{p\gamma}<1$ and occurs at $\varepsilon_\nu\approx 100$\, TeV.
Pion and muon cooling effects dominate in the 4 TeV -- 100 TeV range for WR progenitors and 100 TeV -- 10 PeV range for BSGs. The energies at which these features present themselves at $t=t_{\rm bo}$ do not depend on the choice of proton spectra, as the quantities involved only depend on the progenitor properties. However, as the shape of the proton spectrum gets modified, performing the time integral for fluence can cause visible differences. 

On the right panels of Figure~\ref{NeutrinoSpectra}, we show the time-integrated neutrino flux up to $t=t_{\rm bo}$ for the same configurations. We first discuss the $s=2$ case, which corresponds to the top-right panel. For WR stars, we find that the neutrinos with energies exceeding $10\, {\rm TeV}$ are scarce due to strong IC cooling at breakout times and strong neutrino attenuation at earlier times. The majority of the contribution to the fluence essentially comes from the later times, when the progenitor density is sufficiently small. Since $t_{\rm bo}\sim 60\, {\rm s}$ is rather short, the fluence $\Phi_\nu$ is too small to yield an observable neutrino signal. Neutrino spectra in BSGs can reach the 1 -- 10 PeV range with $L_j \sim 10^{47}-10^{49}\, {\rm erg/s}$ at $t_{\rm bo}$. Furthermore, the neutrino spectrum at injection can extend up to 100 TeV before cooling effects become important. When $P_i=2\, {\rm ms}$ and $B_{\rm dip} = 10^{15}\, {\rm G}$, the comoving temperature is low enough such that $f_{p\gamma}<1$ and neutrino production at late times is suppressed. RSG progenitors have breakout times at $\sim 2000\, {\rm s}$, which allows for $\Gamma_j\sim 10^5$. As mentioned earlier, photons are no longer thermalized after $\sim 400\, {\rm s}$. Hence, neutrino production drops significantly after this time, such that only neutrinos originating from thermal photons at $t\lesssim 400\, {\rm s}$ contribute to the neutrino fluence. We see that the spectral break appears in the $10\; {\rm TeV}-1\; {\rm PeV}$ energy range. In this case, the fluence is largest between 1 PeV and 100 PeV, which accounts for the emission after 100 s. The low-energy tail comes from the superposition of the low-energy tails of emissions at all times, creating a smooth rise in fluence, while the high energy tail end of the neutrino spectrum is the superposition of contributions from late times.

The main difference between the $s=1$ and $s=2$ cases is that, for $s=1$, most of the energy is injected into protons at $\varepsilon_{p}^\prime\sim\varepsilon_{p, {\rm max}}^\prime$. As a result, for a given $\eta_{\rm acc}$, we do not have a large fraction of the low-energy neutrino events. At the same time, all the peaks in $\varepsilon_\nu L_{\varepsilon_\nu}$ and $\varepsilon_\nu \mathcal{E}_{\varepsilon_\nu}$ are shifted to higher energies. As we have more high-energy pions for $s=1$, the total energy deposited into neutrinos is lowered as a result of pion and muon cooling. This affects the cases of WR progenitors the most, where the magnetic fields are stronger during the neutrino emission time as radii and Lorentz factors are small. The strong magnetic fields lower $\varepsilon_{p,{\rm max}}^{\prime}$ and increase synchrotron cooling rates, enhancing the suppression effect above $\varepsilon_\nu\gtrsim $ few TeV. The overall effect is that $\varepsilon_\nu\mathcal{E}_{\varepsilon_\nu}$ has a smooth rise,  resembling a power-law until it reaches the peak fluence, as shown in the bottom-right panel of Figure~\ref{NeutrinoSpectra}.

Both BSG and RSG progenitors are more common than WR stars, making the intrinsic rates of RSG and BSG core-collapse higher than that of WR core-collapse. On the other hand, the occurrence of relativistic jets in these supergiants remain highly uncertain. Such jets are unlikely to correspond to canonical GRBs and thus may not be triggered by current searches for GRB-like transients. As jets originating from these supergiants are harder to detect electromagnetically, their potential as neutrino sources is worth exploring. 

\begin{table}
\centering
\def\arraystretch{1.3}
\caption{Expected number of $\varepsilon_\nu > 1\,{\rm TeV}$ neutrino events detected with IceCube-Gen2 for a source located at $D=100\, {\rm Mpc}$, evaluated for various PNS configurations and density profiles and during the epoch while jet propagates through the progenitor ($t<t_{\rm bo}$). Number of events without (with) brackets correspond to an $\varepsilon_p^{\prime -2}(\varepsilon_p^{\prime -1})$ injected proton spectrum.  
}
\begin{tabular}{|c|c|c|c|}
\hlinewd{0.75pt} 
$(B_{\rm dip}/{\rm G},P_i/{\rm ms})$ &  WR & BSG & RSG \\ \hlinewd{0.25pt} \hline
($10^{15}$, $2$)  & $1.3\times 10^{-2}$ &$4.3 \times 10^{-2}$ & $4.8\times 10^{-2}$ \\  & $(6.9\times 10^{-4})$& $(5.9\times 10^{-3})$ &$(6.3\times 10^{-3})$\\ \hline
$(3\times 10^{15}$, $1.5$) & $5.6\times 10^{-2}$ & $8.9\times 10^{-1}$ & $1.1$\\
& ($1.2 \times 10^{-3})$ & $(5.9\times 10^{-2})$ & $(1.2\times 10^{-1})$ \\ \hline
$(10^{16}$, $1$) & $1.1\times 10^{-1}$ & $14$ & $43$\\
& $(6.5\times 10^{-3})$& $(4.7\times 10^{-1})$ & (3.2) \\
\hlinewd{0.25pt} \hline
\end{tabular}
\label{NuEventsTable}
\end{table}

The neutrino fluxes obtained from Figure~\ref{NeutrinoSpectra} are too low to detect with the current IceCube. This is consistent with the non-observations of neutrinos from GRBs \citep{IceCube:2017amx,IceCube:2018omy,Abbasi_2022} and SNe Ibc \citep{Senno:2017vtd,Esmaili:2018wnv,Chang:2022hqj}. In addition, the recent GRB221009A \footnote{\url{https://gcn.gsfc.nasa.gov/other/221009A.gcn3} \citep{Veres2022}} would constrain the neutrino fluence $E_\nu^2\Phi_\nu\lesssim 3\times 10^{-4}$ erg cm$^{-2}$ for a source at redshift $z=0.15$. For that distance, our WR fluences are $E_\nu^2\Phi_\nu\lesssim 10^{-7}$ erg cm$^{-2}$. 
Likewise, the differential energy of emitted neutrinos $\varepsilon_\nu\mathcal E_{\varepsilon_\nu}\lesssim 10^{49}\, {\rm erg}$ is well below the precursor limits of $\sim 10^{-4}$ erg cm$^{-2}$ from \citet{Abbasi_2022} for GRB180720B and GRB130427A \citep{Gao:2013fra}, at their corresponding redshifts.
We therefore look for detectability in future neutrino detectors. In the case of IceCube-Gen2, we estimate the number of track events from
\begin{equation}
\mathcal{N} = \int dE_\nu A_{\rm eff}(E_\nu) \Phi_{\nu_\mu}(E_\nu),
\label{Ndet}
\end{equation}
where $A_{\rm eff}$ is the neutrino effective area and $\Phi_{\nu_\mu}$ is the muon neutrino flux. To get $A_{\rm eff}$ we use the effective area in \cite{Stettner:2019tok} and scale it by a factor of $10^{2/3}$ to account for IceCube-Gen2's detector size.

In Table \ref{NuEventsTable}, we list the expected number of neutrino events that can be detected with IceCube-Gen2, for a source located at distance $D=100\,{\rm Mpc}$, for both $s=1$ and $s=2$. We take $E_{\nu,{\rm min}} = 1\, {\rm TeV}$ as the minimum detectable neutrino energy. We first analyse $s=2$, which is our  proton injection spectrum which leads to a larger number of neutrino events. Even in the most optimistic PNS configuration with strong field and rapid rotation, we find that WR stars has a negligible neutrino signal. 
For the BSG progenitor, only the most optimistic scenario with $(B_{\rm dip},P_i)=(10^{16}\, {\rm G}, 1\, {\rm ms})$ yields a few detectable neutrino events. 
The RSG progenitor presents the most promising scenario with several neutrino events detectable above $10\, {\rm TeV}$, resulting from the typically large jet Lorentz factors for times close to $t_{\rm bo}$. In this case, we can get up to a few tens of events for optimistic central engine configurations. For a harder spectral index $s=1$, we see that the expected number of neutrino events drops by roughly an order of magnitude. This is explained by the reduction of lower-energy neutrinos. The neutrino-nucleon cross section and $A_{\rm eff}$ do not increase sufficiently fast with energy to compensate for this lack of low-energy neutrino flux, which results in fewer detectable events. 
However, one should keep in mind that the results for $s=1$ are sensitive to $\varepsilon_{p, {\rm max}}^\prime$, which can be lowered for larger values of $\eta_{\rm acc}$.

\section{Discussion \& Implications}
\label{Sec6}
\subsection{Impacts of the progenitor}
Relativistic jets can be launched from highly-magnetized rapidly-rotating PNS that are formed shortly after stellar core collapse. 
The Poynting flux is initially stored close to the PNS and gradually converted into jet kinetic energy post launch.  
This leads to an increase in $\sigma_0$ over time which subsequently facilitates dissipation (see Figure~\ref{sigma0_Edot_t}). More energetic jets emerge from PNSs with a combination of stronger field and rapid rotation, coupled with smaller jet opening angle.  
In this study, we analytically investigated the properties of such magnetized outflows as they propagate through their stellar progenitors, in particular considering BSGs and RSGs as well as stars with stripped He envelopes such as WRs. 

The jet collimation occurs due to the formation of oblique shock close to jet base and depends on the strength of jet-cocoon interactions. For $B_{\rm dip} \sim 10^{15}-10^{16}\, {\rm G}$ and $P_i \sim 1-2\, {\rm ms}$, the jet remains uncollimated before breakout as the jet pressure exceeds the cocoon pressure, $P_j \gtrsim P_c$ for $t \lesssim t_{\rm bo}$. As the jet-head attains relativistic velocities at later times inside more dense stellar media, it is easier for jets with larger $\theta_j$ to be collimated inside WR stars. By comparison, the cocoon remains sub-relativistic throughout the cooling phase.

Unlike hydrodynamic jets, magnetized jets have a narrower jet cross-section and encounter less stellar material prior to breakout. Therefore, they propagate much faster with a shorter $t_{\rm bo}$ and dissipate considerably less energy to the cocoon $E_c(t_{\rm bo})$ while crossing the stellar envelope. As expected, the relativistic jets originating from PNS with stronger fields and rapid rotation rates deposit more energy into their surrounding cocoon. The deposited energy at $t_{\rm bo}$ tends to be larger for BSGs ($E_c \sim 10^{46}-10^{50}\, {\rm erg}$) and RSGs ($E_c \sim 10^{47}-10^{51}\, {\rm erg}$) progenitors, with significantly longer $t_{\rm bo}$, compared to their WR counterparts ($E_c \sim 10^{45}-10^{49}\, {\rm erg}$). 

Jets can get choked within the stellar envelope if the PNS stops at $t < t_{\rm th}$ before the jet exits star or if 
$\tilde{E}_{\rm j,iso} = \int_0^{t_{\rm bo}}L_{j,\rm iso}dt$ is smaller than the minimum energy $\tilde{E}_{\rm ej}$ required to push through the stellar envelope. Magnetized jets with smaller $\theta_j$ are more likely to break out 
as $\tilde{E}_{\rm j,iso}(t_{\rm bo})/\tilde{E}_{\rm ej} \propto \theta_j^{-4}$.  
The rotation rate $P_i$ does not affect the breakout criterion, whereas relativistic outflows from PNS with weaker fields $B_{\rm dip} \lesssim 3\times10^{15}\, {\rm G}$ can get choked inside the stellar envelope of WR stars. However, jets with $10^{15} \lesssim B_{\rm dip}/{\rm G} \lesssim 3\times10^{16}$ and $1 \lesssim P_i/{\rm ms} \lesssim 5$ tend to break out from their less dense BSG and RSG counterparts.  

Magnetic fields can stabilise the jet by inhibiting the growth of local instabilities along the jet-cocoon boundary. While wider and/or low-power jets require stronger fields for stabilisation, narrower and/or high-power jets tend to be more stable. 
Current-driven kink instabilities can arise in magnetized jets which can render the jet structure globally unstable. We find that magnetized jets with $10^{15} \lesssim B_{\rm dip}/{\rm G} \lesssim 3\times10^{16}$ and $1 \lesssim P_i/{\rm ms} \lesssim 5$ are always stable against kink instability as $t_{\rm kink}/t_{\rm dyn} \gg 1$ throughout their propagation in the stellar envelope. Furthermore, jets with wider opening angles are more stable as $t_{\rm kink} \propto r_j^2$.

\subsection{Model assumptions}
This study makes several assumptions to simplify the analytical modeling of  magnetized jet propagation in the stellar ejecta of GRB progenitors. Firstly, we assume an axisymmetric jet that expands into uniform cold stellar medium i.e. the external pressure is negligibly small and does not influence dynamical evolution of the system. The jet is launched with an opening angle that does not vary with time.  
We approximate the cocoon pressure as being uniform and assume that the jet material does not lose energy due to the work done against cocoon pressure as its propagates from the injection point to the jet-head. This is justified as the jet injection angle is fixed and small, and that the stellar envelope does not expand with time \citep{Bromberg2011}. Although we assume a fixed PNS mass for this study, central engine properties such as its size, magnetic field and/or rotation rate should correlate with properties of the stellar envelope. 
Lastly, $B_{\rm dip}$ can be dynamically amplified over the cooling phase due to differential rotation of the PNS outer layers, an effect that we do not consider here.

\subsection{Neutrinos}
If protons are accelerated in the magnetized jet through magnetic dissipation, they must interact with ambient photons escaping from the termination shock to generate pions. This subsequently leads to the production of high-energy neutrinos with $\varepsilon_{\nu} \gtrsim 1\, {\rm TeV}$ via $p\gamma$ interactions. The signatures of neutrino oscillation have been of much interest. The oscillations of these neutrinos in the context of precursor or orphan neutrinos have been studied using both analytical and numerical methods \citep{Mena2007,Sahu2010,Razz2010,Xiao2015,Carpio2020,Abbar:2022hgh}. 

PNSs with stronger magnetic fields and rapid rotation rates potentially have larger intrinsic powers.
However, in WR stars the high-energy neutrino flux is largely suppressed. The resulting neutrino fluence is too small for the detection even with IceCube-Gen2. This is consistent with previous conclusions \citep[e.g.,][]{MuraseIoka2013}. Canonical GRB jets propagating in a WR star are unlikely to be efficient sources of high-energy neutrinos but quasi-thermal neutrinos may still be detectable \citep{GaoMeszaros2012,MKM2013,Kashiyama2013}.

For the BSG and RSG progenitors with larger radii, the observed neutrino fluence is a result of late time emission when attenuation is negligible. Neutrinos from BSG sources are only detectable for energetic outflows with strong $B_{\rm dip}\gtrsim 3\times10^{15}\, {\rm G}$, rapid $P_i \lesssim 1.5\, {\rm ms}$ and $s\approx 2$, unless the source distance is significantly smaller than 100 Mpc. In the case of $s=1$, BSG leads to no detectable neutrinos if the source distance is 100 Mpc. The high luminosities and large Lorentz factors for magnetized jets in RSG progenitors present the most promising scenario with $s=2$
for the detection of $\mathcal{N} > {\rm few}\ \times 10$ events above 10 TeV with IceCube-Gen2. For a harder spectrum with $s=1$, $\mathcal{N}$ is marginal but we could still detect $\mathcal{N}\sim 3$ events for the most energetic configuration.

As noted above, the jet may be accelerated to achieve a large Lorentz factor inside the expanding magnetized bubble. Once the jet leaves the bubble (before the breakout from the star), the mixing with the ambient cocoon may occur. In this case, one should consider radiation constraints \citep{MuraseIoka2013} for a hydrodynamic jet with $z_{\rm cs}>R_w$.

\subsection{Cosmic-rays}
The relativistic winds studied in this paper are promising ultra-high energy cosmic ray (UHECR) sources as their environments consist of primarily heavy nuclei \citep{Murase:2006mm,Murase2008,Metzger2011b,Horiuchi2012,MBh2021,Ekanger2022}. Recent measurements by the Pierre Auger Observatory \citep{PAO2015} indicate that UHECR composition at high energies is primarily dominated by heavier nuclei \citep{Abraham2010,Taylor2011,Abbasi2018,Batista2019}. Since relativistic outflows in rapidly rotating magnetars can also power GRBs, they can simultaneously synthesize and accelerate heavy nuclei to ultrahigh energies, making them intriguing nuclei UHECR sources. Therefore, it is important to consider the effect of jet propagation and subsequent mixing with its surrounding medium on the composition of outflows and UHECR \citep{GN21,HI2021}. For our analysis, we have considered PNS with mass $M_{\rm ns}=1.4\, M_{\odot}$, dipole magnetic fields $3\times10^{14}\, {\rm G} \lesssim B_{\rm dip} \lesssim 3\times10^{16}\, {\rm G}$, rotation periods $1\, {\rm ms} \lesssim P_i \lesssim 5\, {\rm ms}$ and obliquity angle $\chi = \pi/2$. We find that before breakout, any nuclei synthesized in the outflow will be disintegrated due to the large photon density in the outflow, implying that UHECRs need to be sourced at later epochs (see, e.g., \citealt{MBh2021,Ekanger2022}).

\section{Summary \& Conclusions}
\label{Sec7}
Relativistic jets powered by strongly magnetized and rapidly rotating protomagnetars may be relevant for GRBs, and have been investigated as potential sources of UHECRs and very high energy neutrinos.
In this work, we used a semi-analytical model for protomagnetar spin-down to investigate the role of central engine properties (namely $B_{\rm dip}$, $P_i$ and $\theta_j$) on the dynamical evolution of the jet-cocoon system, its interaction with the surrounding stellar material, and the production of high-energy neutrinos. 
The time evolution of the jet is determined by the time-dependent luminosity and outflow magnetization which are obtained from the protomagnetar spin-down.  For a broad range of protomagnetar parameters and potential GRB progenitors, we argued that magnetized jets can be stable against current-driven kink instabilities such that they are uncollimated when they break out (especially, for $B_{\rm dip} \gtrsim 10^{15}\, {\rm G}$ and $P_i \lesssim 2\, {\rm ms}$). While relativistic jets break out for most protomagnetar configurations, the breakout time is longer for BSG and RSG progenitors and the jet can therefore deposit considerably more energy into the cocoon.   

Late-time neutrino emission contributes the most towards the detectability of precursor neutrino signals, as neutrino absorption from the source environment is the weakest at that stage. We find that the expected fluxes for magnetized jets powered by protomagnetars are below the IceCube detection range, but may be observed in IceCube-Gen2 if the external material is sufficiently extended. For WR stars, there is little observable signal for a source distance of 100 Mpc and neutrinos typically get absorbed by their dense interiors; for BSGs, when $B_{\rm dip} \gtrsim 3\times10^{15}\, {\rm G}$ and $P_i \lesssim 1.5\,{\rm ms}$, which correspond to the most energetic jets, one might detect $\sim 15$ neutrino events with IceCube-Gen2 for $s=2$, but fewer events if $s=1$. Magnetized outflows from RSG progenitors at distances up to 100 Mpc could yield as many as $\sim 40$ detectable neutrino events with IceCube-Gen2 for $s=2$ and $\sim 3$ events for $s=1$. The precursor neutrino signatures from BSG and RSG progenitors are one of the few ways of observing jets launched in their core collapses which are otherwise hard to directly detect electromagnetically. Thus, searches for neutrinos coincident with possible nearby transients from the collapse of RSGs and BSGs may provide an unique opportunity to study the presence of jets using the multimessenger approach.

\section*{Acknowledgements}
We thank Nick Ekanger, Peter Mészáros and David Radice for useful discussions. We thank Ke Fang, Kunihito Ioka, Mainak Mukhopadhyay, Daichi Tsuna and Bing Theodore Zhang for carefully reading the manuscript and providing insightful comments, and Chie Kobayashi for help with figures. 
MB acknowledges support from the Eberly Research Fellowship at the Pennsylvania State University. 
J.C. is supported by the NSF Grant No. AST-1908689 and No. AST-2108466.
The work of K.M. is supported by the NSF Grant No.~AST-1908689, No.~AST-2108466 and No.~AST-2108467, and KAKENHI No.~20H01901 and No.~20H05852. 
The work of SH is supported by the U.S.~Department of Energy Office of Science under award number DE-SC0020262, NSF Grant No.~AST1908960 and No.~PHY-1914409 and No.~PHY-2209420, and JSPS KAKENHI Grant Number JP22K03630. This work was supported by World Premier International Research Center Initiative (WPI Initiative), MEXT, Japan. 

\section*{Note added}
Recently, \citet{Guarini2022} performed numerical simulations of collapsar jets to study neutrino production. 
Our work is inherently distinct from their work. 
We considered large values of $\sigma_0$ which are difficult to study with current magnetohydrodynamic simulations, and explored different progenitors such as BSGs and RSGs which are much more extended than their WR counterparts. Some of the descriptions of \cite{Carpio2022} presented in their ``Note Added'' are inaccurate. In particular, the explored parameter space for jets and progenitors in these works are different, and we considered photomeson production and magnetic reconnections as the relevant mechanisms. 

\section*{Data availability}
The data underlying this article will be shared on reasonable request to the corresponding author.

\bibliography{main}

\appendix 

\section{Notation Table}
Here we include a table listing the symbols that we use along with their physical description. We also mention the first equation or section where they are used. `Text' is used to denote a symbol that comes from equation within the text.

\begin{table*}
\begin{center}
\caption{List of symbols used in this work along with their physical description and where (equation or section) they are first used. We write `Text' for the symbols that come from equations within the text.}
\label{Table1}
\bgroup
\def\arraystretch{1.0}
\begin{tabular}{|c|c|c|}
\hlinewd{0.25pt} \hline
\centering
\textbf{Symbol} & \textbf{Description} & \textbf{Equation/Section} \\ \hline \hlinewd{0.25pt}
& \textbf{Central engine parameters} & \\ \hlinewd{0.75pt}
$L_{\nu}$, $\epsilon_{\nu}$ & Neutrino luminosity, mean energy & (\ref{Mdot})/\S\ref{S2.1}\\
$\dot{M}$ & Mass loss rate due to neutrino-heated wind & (\ref{Mdot})/\S\ref{S2.1} \\
$M_{\rm NS}$, $R_{\rm NS}$ & Protomagnetar mass, radius & (\ref{Mdot})/\S\ref{S2.1} \\
$\chi$ & Magnetic obliquity angle & Text/\S\ref{S2.1} \\
$\Omega=2\pi/P$ & PNS angular velocity for spin period $P$ & Text/\S\ref{S2.1} \\ 
$\eta_s$ & Stretch factor for the neutrino quantities & Text/\S\ref{S2.1} \\
$\phi_B$ & Magnetic flux due to surface dipole field $B_{\rm dip}$ & Text/\S\ref{S2.1} \\
$R_{\rm mag}$ & Magnetic dissipation radius & (\ref{R_mag})/\S\ref{S2.1} \\
$\dot{E}_{\rm kin}$, $\dot{E}_{\rm mag}$ & Kinetic, magnetic wind luminosity & Text/\S\ref{S2.1} \\
$J$ & Angular momentum of rotating PNS & Text/\S\ref{S2.1} \\
\hlinewd{0.75pt}
& \textbf{Jet/outflow parameters} & \\ \hlinewd{0.75pt}
$f_{\rm open}$ & Fraction of PNS surface threaded by open field lines & (\ref{Mdot})/\S\ref{S2.1} \\
$f_{\rm cent}$ & Enhancement to $\dot{M}$ from magnetocentrifugal effect & (\ref{Mdot})/\S\ref{S2.1} \\
$C_{\rm es}$ & Heating correction for inelastic neutrino-electron scatterings & (\ref{Mdot})/\S\ref{S2.1} \\
$\Gamma_j$ & Jet Lorentz factor & (\ref{Gamma_j})/\S\ref{S2.1} \\ 
$\sigma_0$ & magnetization of the outflow & Text/\S\ref{S2.1} \\
$L_j$ & Jet luminosity & Text/\S\ref{S2.1}\\ 
$\theta_j$ & Jet opening angle & Text/\S\ref{S2.1}\\
$h_{j/a}$ & Specific enthalpy of the jet/ambient medium & (\ref{ramP_bal})/\S\ref{S3.1}\\
$\Gamma_{jh}$ & Relative Lorentz factor between jet and jet-head & (\ref{ramP_bal})/\S\ref{S3.1}\\ 
$P_j$, $r_j$ & Jet pressure, cross-sectional radius & Text/\S\ref{S3.1}\\
$\Gamma_{h}$ & Lorentz factor of the jet-head & (\ref{ramP_bal})/\S\ref{S3.1}\\ 
$\tilde{L}$ & Jet energy density/ambient energy density & (\ref{Ltilde})/\S\ref{S3.1}\\
$\tilde{P}$ & Jet pressure/ambient pressure & (\ref{Ltilde})/\S\ref{S3.1}\\
$r_h$ & Jet-head position & (\ref{P_c})/\S\ref{S3.2}\\
$r_s$ & Cylindrical radius of the collimation shock & \S\ref{S3.3}\\ 
$z_{\rm cs}$ & Converging position of the collimation shock & (\ref{zhat})/\S\ref{S3.3}\\ \hlinewd{0.75pt}
& \textbf{Ejecta parameters} & \\ \hlinewd{0.75pt}
$\rho_a(r)$ & Density profile of the external medium & Text/\S\ref{S2.3}\\
$\alpha$ & Density profile power-law index & Text/\S\ref{S2.3}\\
$M_*$, $R_*$ & Mass, radius of the stellar progenitor & Text/\S\ref{S2.3}\\
$Z$ & Metallicity corresponding to stellar progenitor composition & Text/\S\ref{S2.3}\\
$P_{\rm ext}$ & Pressure of external medium & (\ref{ramP_bal})/\S\ref{S3.1}\\ \hlinewd{0.75pt} 
& \textbf{Cocoon parameters} & \\ \hlinewd{0.75pt}
$E_c$ & Energy deposited into the cocoon by jet & Text/\S\ref{S3.2}\\
$\eta$ & Fraction of jet energy deposited into cocoon & Text/\S\ref{S3.2}\\
$P_{c}$, $V_c$, $r_c$ & Cocoon pressure, cylindrical volume, cross-sectional radius & (\ref{P_c})/\S\ref{S3.2}\\
$\Gamma_c$ & Lorentz factor of the cocoon & (\ref{P_c})/\S\ref{S3.2}\\ 
$\overline{\rho}_a$ & Mean density of the surrounding medium & (\ref{beta_c})/\S\ref{S3.2}\\ \hlinewd{0.75pt}
& \textbf{System timescales} & \\ \hlinewd{0.75pt}
$t_{\rm bo,hyd}$ & Breakout time for hydrodynamic jets  & (\ref{tbo_hyd})/App\ref{App2}\\
$t_{\rm bo,mag}$ & Breakout time for magnetized jets & (\ref{tbo_mag})/App\ref{App2} \\
$t_{\rm eng/th}$ & Central engine/threshold activity time & Text/\S\ref{S4.2}\\
$t_{\rm kink}$ & Timescale for kink instability to develop & (\ref{tkink})/\S\ref{S4.3}\\
$t_{\rm dyn}$ & Outflow dynamical/expansion timescale & Text/\S\ref{S4.3}\\
\hlinewd{0.75pt}
& \textbf{CR and neutrino spectrum parameters} & \\ \hlinewd{0.75pt}
$dN/d\varepsilon$ & Proton/neutrino spectrum & Text/\S \ref{Sec5} \\
$\epsilon_B$, $\epsilon_p$ & Luminosity fraction converted to magnetic field energy, proton energy & Text/\S \ref{Sec5} \\ 
$f_{p\gamma}$ & Effective optical depth for $p\gamma$ interactions & Text/\S \ref{Sec5} \\
$f_{\rm sup}$ & Spectrum suppression factors & Text/\S \ref{Sec5}\\
$\phi_{\nu}$, $\Phi_{\nu}$ & Neutrino flux, fluence & Text/\S \ref{Sec5}\\ 
$D$ & Neutrino source distance & Text/\S \ref{Sec5}\\
$\mathcal{N}$ & Detected events with IceCube-Gen2 & (\ref{Ndet})/\S \ref{Sec5}\\
$E_{\nu,\rm min/max}$ & Minimum/maximum neutrino energy & (\ref{Ndet})/\S \ref{Sec5}\\
$\sigma_{\nu N}(E_{\nu})$ & Neutrino-nucleon cross section & (\ref{Ndet})/\S \ref{Sec5}\\
$n_{\gamma}$, $T_{\gamma}$ & Photon number density, temperature & Text/\S \ref{Sec5}\\
$f_{\rm esc}$ & Photon escape fraction from termination shock & Text/\S \ref{Sec5}\\
$n_{\rm sj}$ & Number density of particles in shocked jet & Text/\S \ref{Sec5}\\
$\sigma_{p\gamma}$ & Cross section for $p\gamma$ interactions & \eqref{fpgamma}/\S \ref{Sec5}\\\
$L_{\varepsilon_\nu}$ & $\varepsilon_\nu dN_\nu /d\varepsilon_\nu$  & \eqref{fpgamma}/\S \ref{Sec5}\\
\hlinewd{0.75pt}
\end{tabular}
\egroup
\end{center}
\end{table*}

\section{Jet breakout time}
\label{App2}
Prior to breakout, the jet collides with the stellar envelope to generate a reverse shock. The shocked material from the jet and the envelope move sideways from the jet head to form a cocoon. At the expense of this shocked matter, the jet head moves outwards and drills a hole into the stellar envelope. This is called the jet breakout. The high pressure cocoon confines the jet through a CS before the jet breaks out. After jet breakout, the jet expands into the circumstellar medium which is assumed to be very dilute. 
Unlike hydrodynamic jets that typically cross the star at sub-relativistic velocities, Poynting flux dominated jets have a narrower jet head and therefore encounter less resistance by the stellar material. Consequently, these magnetized jets move much faster with a shorter $t_{\rm bo}$ and dissipate much less energy while crossing the stellar envelope.

\citet{Bromberg2015} derived the jet breakout time assuming canonical values for the stellar mass $M_*=15\, M_{\odot}$, stellar radius $R_*=4\, R_{\odot}$ and a power-law density profile $\rho_* \propto r^{-2.5}$. For hydrodynamic jets, the breakout time is 
\bea \label{tbo_hyd}
t_{\rm bo,hyd} = (6.5\, {\rm s})R_{*,4R_{\odot}}\left[\left(\frac{L_j}{L_{\rm rel}}\right)^{-2/3} + \left(\frac{L_j}{L_{\rm rel}}\right)^{-2/5}\right]^{1/2}
\eea
Here $L_{\rm rel} \sim (1.6\times10^{49}\, {\rm erg\, s^{-1}})R_{*,4R_{\odot}}^{-1} M_{*,15M_{\odot}}\theta_{0.84}^4$ is the transition luminosity between a non-relativistic breakout time and a relativistic one. As a Poynting flux dominated jet becomes relativistic deep within the star, the corresponding breakout time is \bea
t_{\rm bo,mag} = (9.2\, {\rm s})R_{*,4R_{\odot}}\left(1 + 0.11 L_{j,49}^{-1/3} R_{L,7}^{2/3} M_{*,15M_{\odot}}^{1/3} R_{*,4R_{\odot}}^{-1}\right)
\label{tbo_mag}
\eea
where $R_{*,4R_{\odot}} = R_*/4R_{\odot}$ and $M_{*,15M_{\odot}} = M_*/15M_{\odot}$. The breakout time for a magnetized jet is significantly smaller compared to the breakout time for a hydrodynamic jet with similar luminosity and progenitor star parameters. 
From equation~(\ref{tbo_mag}), the breakout times for the $(15.7M_{\odot},5.15R_{\odot})$ WR, $(15.9M_{\odot},52R_{\odot})$ BSG and $(11.9M_{\odot},875R_{\odot})$ RSG progenitors are $14.6\, {\rm s}$, $122.2\, {\rm s}$ and $2023\, {\rm s}$, respectively.  
As $t_{\rm bo,mag}$ does not have a strong dependence on $L_j$, it is primarily decided by the stellar radius $R_*$ instead of the specific $(B_{\rm dip},P_i)$ configuration considered.

\section{Criteria for jet collimation before breakout}\label{App3}

\begin{figure*} 
\includegraphics[width=0.32\textwidth]{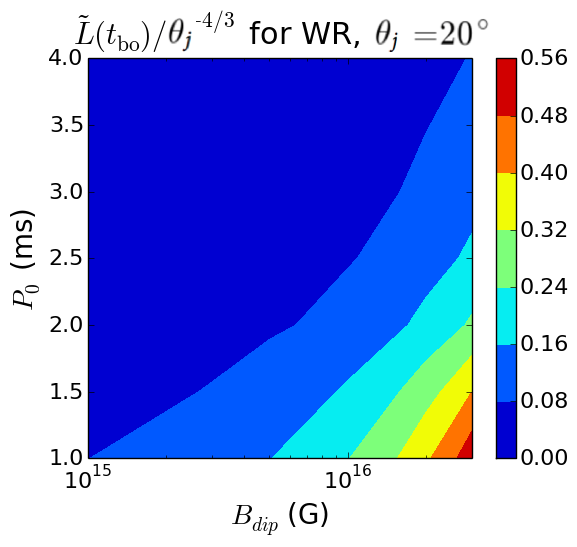} 
\includegraphics[width=0.32\textwidth]{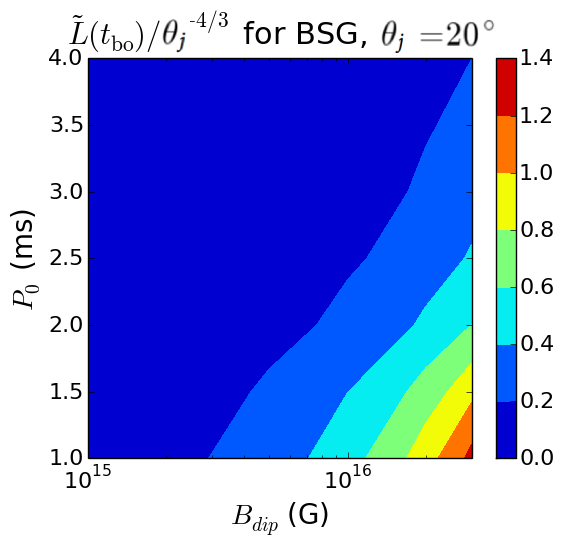}
\includegraphics[width=0.32\textwidth]{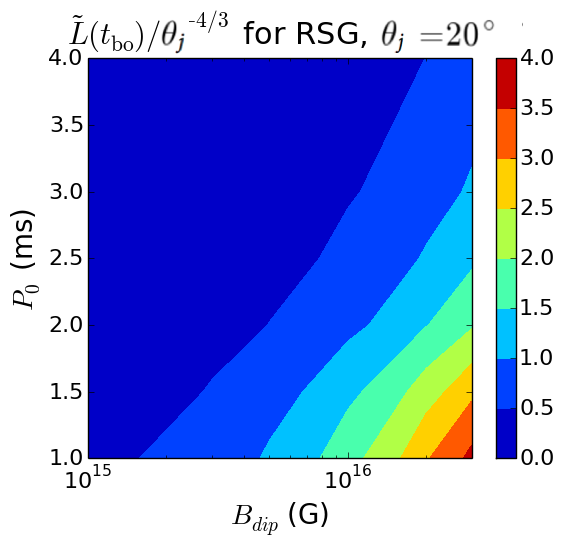}
\vspace{-0.3cm}
\caption{Contour plots for $\tilde{L}(t_{\rm bo})/\theta_j^{-4/3}$ shown in $B_{\rm dip}-P_i$ plane, where $\tilde{L}$ is given by equation (\ref{Ltilde}) and jet-opening angle $\theta_j = 20^{\circ}$. A relativistic jet gets collimated before it breaks out of the stellar envelope provided that $\tilde{L}(t_{\rm bo}) \lesssim \theta_j^{-4/3}$ and $P_j \lesssim P_c$. For a given stellar density profile, it is generally easier to collimate jets with larger opening angles $\theta_j$ as their isotropic-equivalent luminosities are relatively smaller. On the other hand, magnetized jets are more likely to remain uncollimated (with $\tilde{L}(t_{\rm bo}) \gtrsim \theta_j^{-4/3}$, for similar values of $B_{\rm dip}$ and $P_i$) if the external medium is less dense, for e.g., BSG and RSG progenitors.
} 
\label{Ltilde_contours}
\end{figure*}

\begin{figure*} 
\includegraphics[width=0.32\textwidth]{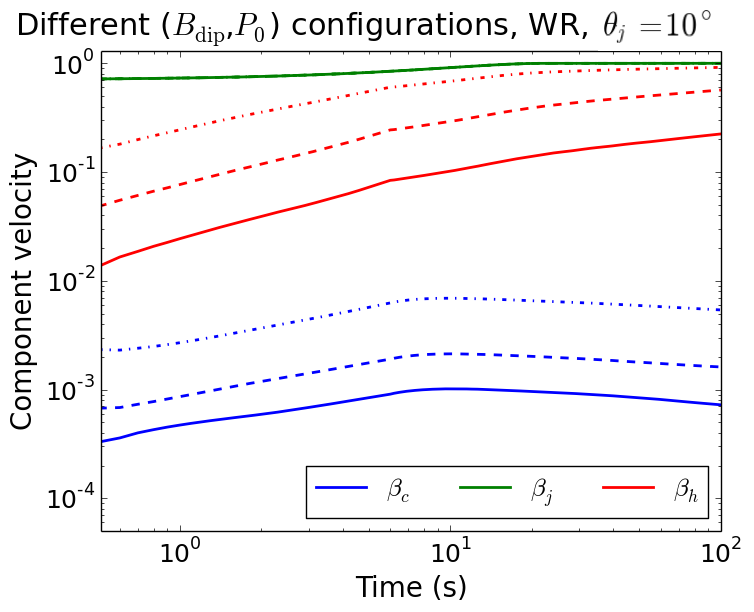} 
\includegraphics[width=0.32\textwidth]{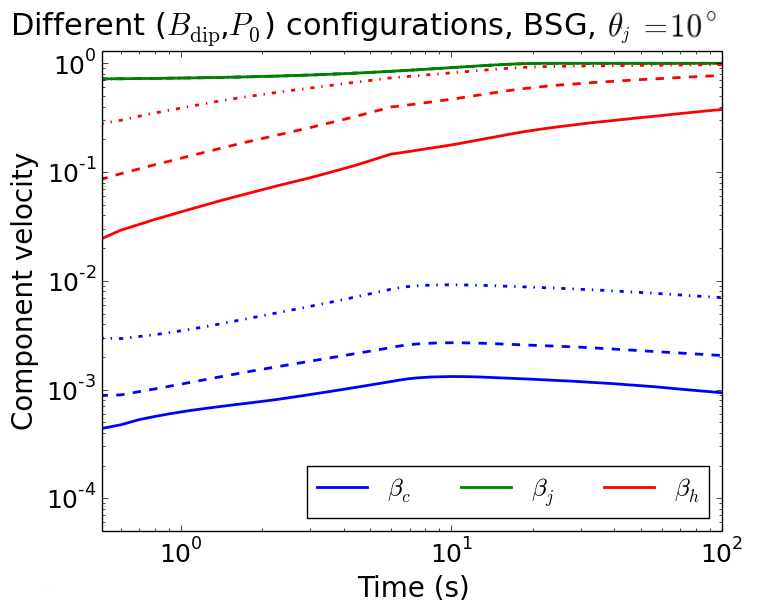}
\includegraphics[width=0.32\textwidth]{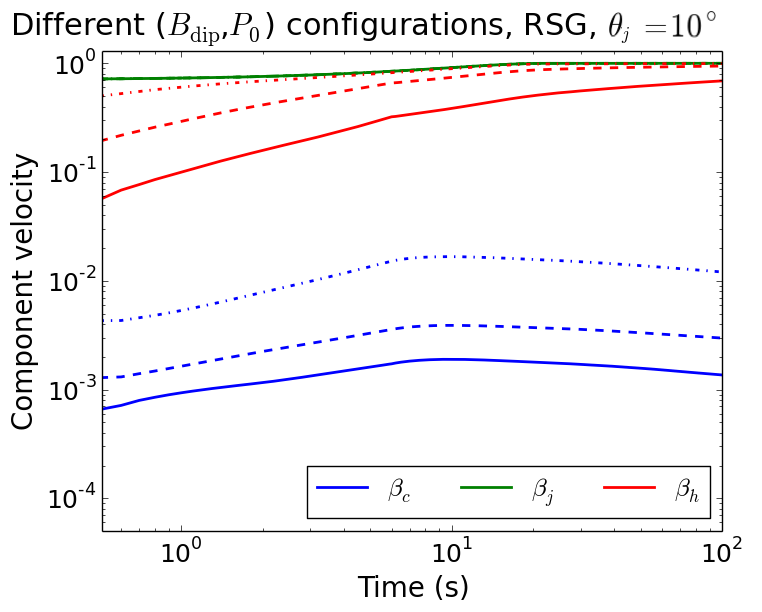}
\vspace{-0.3cm}
\caption{Variation of cocoon, jet and jet-head velocities with time are shown for three PNS configurations with $(B_{\rm dip},P_i)=(10^{15}\, {\rm G},2\, {\rm ms})$ [solid curves], $(3\times10^{15}\, {\rm G},1.5\, {\rm ms})$ [dashed curves] and $(10^{16}\, {\rm G},1\, {\rm ms})$ [dot-dashed curves]. The left, center and right panels are shown for WR, BSG and RSG progenitor, respectively.  
While both $\beta_c$ and $\beta_h$ tend to increase with an increase in the progenitor radius $R_*$, the effect on the jet-head velocity is more pronounced. 
} 
\label{comp_velocity}
\end{figure*}

\begin{figure*} 
\includegraphics[width=0.325\textwidth]{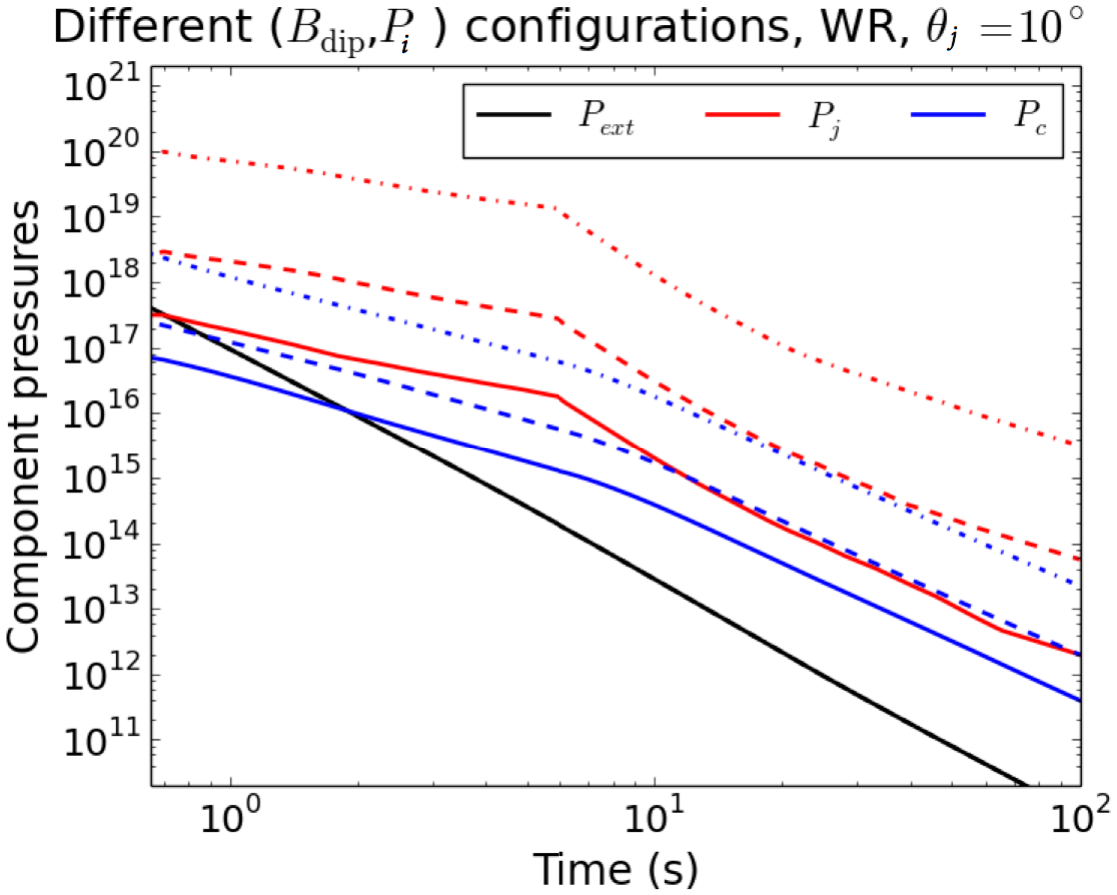}
\includegraphics[width=0.325\textwidth]{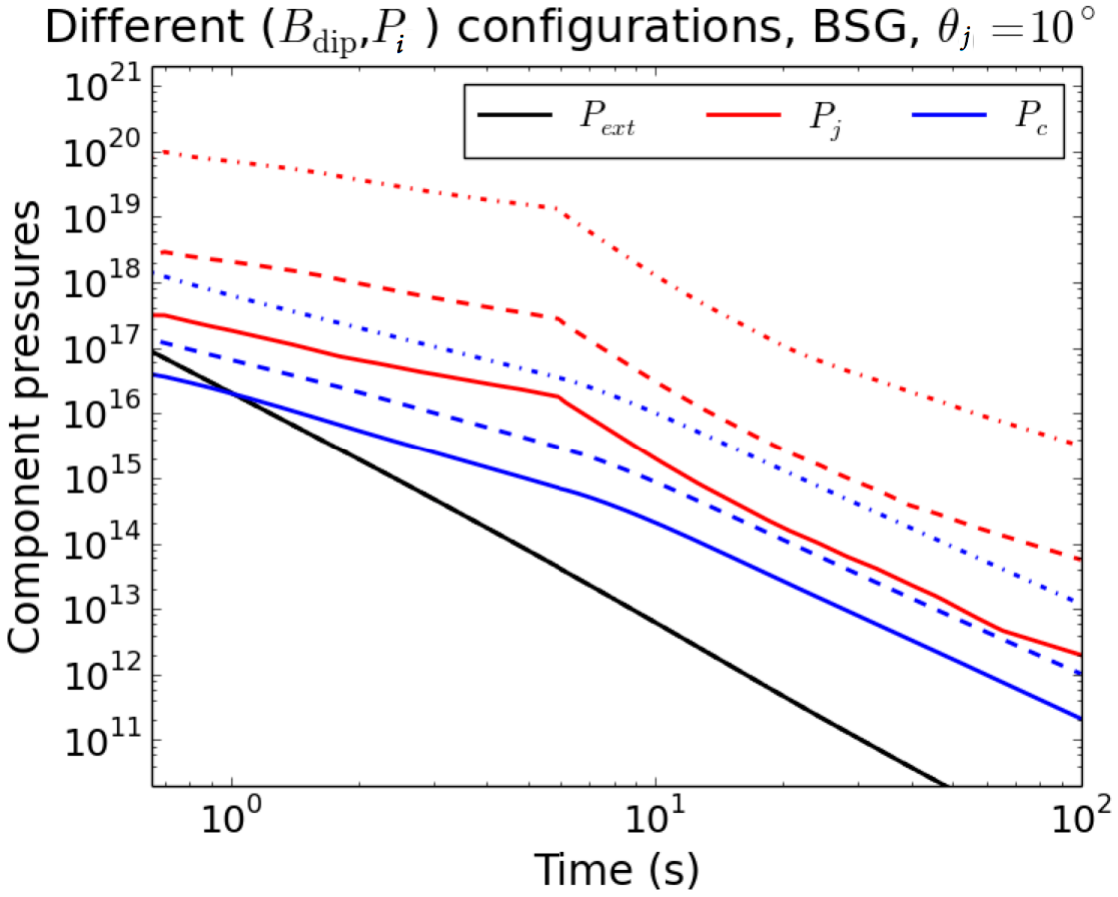}
\includegraphics[width=0.325\textwidth]{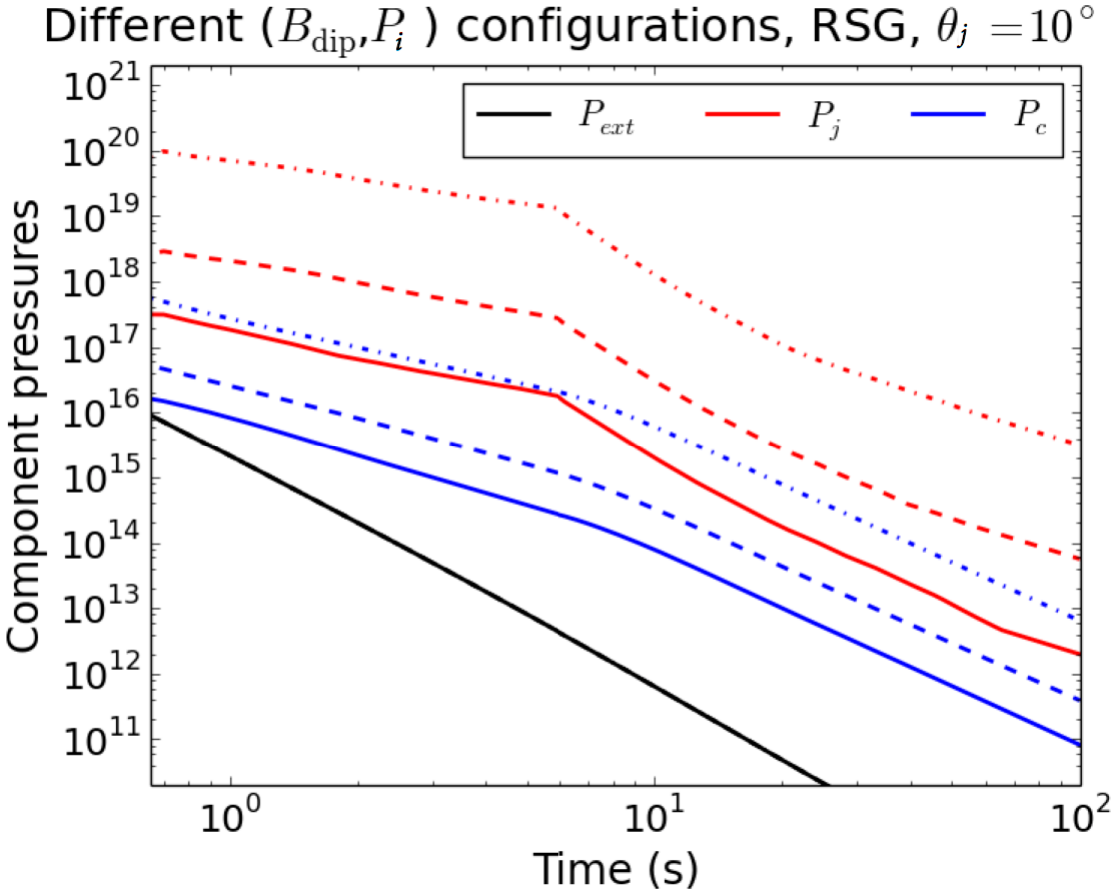}
\vspace{-0.3cm}
\caption{Time evolution of the external medium pressure, jet pressure and cocoon pressure are shown for same PNS configurations and stellar density profiles as in Figure~\ref{comp_velocity}. Jet pressure scales with the initial opening angle as $P_j \sim L_j/r_j^2 \sim \theta_j^{-2}$ once the magnetized outflow becomes relativistic,  
increasing for smaller $\theta_j$ as jet energy density rises. While $P_c$ and $P_{\rm ext}$ are independent of $\theta_j$, $P_c$ is marginally higher for WR stars with smaller $R_*$ compared to BSG and RSG progenitors, as density of the external medium is relatively larger.} 
\label{comp_pressure}
\end{figure*}

Magnetized jets that arise from PNS central engines are either collimated or uncollimated based on the jet energy density ($L_j/\pi r_j^2 c)$ and its initial opening angle $\theta_j$. For sufficiently high $P_c$, jet collimation occurs which reduces the jet-head cross section and accelerates its propagation through the dense stellar medium. There is an upper limit to $\beta_h$ above which $P_c$ becomes too low to effectively collimate the jet and therefore the jet can transition to an uncollimated state before it finally breaks out of the stellar envelope. 

The critical jet parameter  $\tilde{L} = L_j/(\pi r_j^2 \rho_a c^3)$ varies as the jet propagates, depending on the stellar density profile and behaviour of the jet’s cross section. $\tilde{L}$, together with $\theta_j$, generally decides whether the jet is collimated by the cocoon or not. It is easier to collimate the outflow inside the star for sufficiently low-power jets ($\tilde{L}(t_{\rm bo}) \lesssim \theta_j^{-4/3}$), as they become slow and cylindrical. Contrastingly, when $\tilde{L}(t_{\rm bo}) \gg \theta_j^{-4/3}$, the cocoon pressure is generally too weak to laterally compress the jet and it remains conical in geometry. 

Figure \ref{Ltilde_contours} shows the  $\tilde{L}/\theta_j^{-4/3}$ contours in $B_{\rm dip}-P_i$ plane, evaluated at the time of jet breakout. The results are shown for WR, BSG and RSG progenitors, and for jet opening angle $\theta_j = 20^{\circ}$. The relativistic jet is collimated by the surrounding medium before breakout provided $\tilde{L}/\theta_j^{-4/3} \lesssim 1$, for a given $(B_{\rm dip}, P_i)$ configuration and stellar density profile. Jets with a wider opening angle are somewhat easier to get collimated by the cocoon due to their marginally smaller energy densities for a similar central engine configuration. Irrespective of the progenitor density profile, jets that arise from PNS with stronger fields and rapid rotation rates are more likely to be uncollimated at the breakout time due to their large energy density. However, for the WR star with $M_*=15.7\, M_{\odot}$ and $R_*=5.15\, R_{\odot}$, we find that $\tilde{L}/\theta_j^{-4/3} \lesssim 1$ across the entire $B_{\rm dip}-P_i$ considered, indicating that most jets within such progenitors can pontetially be collimated before breakout if $P_{\rm jet} \lesssim P_c$. 

A fast head is essential for successful jet breakout from the confining medium during the lifetime of the central engine. Without strong collimation, the jet head would remain buried deep in the medium when the engine ceases to be active, and jets would fail to break out and produce luminous GRBs. Initially, the internal pressure of the jet is so large that it expands freely until the collimation point where $P_j = P_c$. After this point, the jet is collimated by the $P_c$ with the transition to collimated state accompanied by oscillations in $r_j$ around its equilibrium value. 

Figure~\ref{comp_velocity} shows the evolution of velocities for different system components, namely the cocoon, jet and the jet-head for $\theta_j=10^\circ$. The results are shown for three PNS configurations with $(B_{\rm dip},P_i)=(10^{15}\, {\rm G},2\, {\rm ms})$, $(3\times10^{15}\, {\rm G},1.5\, {\rm ms})$ and $(10^{16}\, {\rm G},1\, {\rm ms})$. The velocity profiles for the WR, BSG and RSG progenitors are shown in the left, center and right panels, respectively. 
We find that the jet opening angle does not affect the component velocities of the jet-cocoon system for $\theta_j \sim 5-20^{\circ}$. While $\beta_c$ remains sub-relativistic for $t \sim t_{\rm KH}$, the jet-head powered by $\sigma_0$ can attain relativistic velocities at later times $t \gtrsim {\rm few}\ 10\ {\rm s}$. For a given progenitor density profile, magnetized jets with a combination of stronger fields and rapid rotation rates achieve higher $\beta_c$ and $\beta_h$. While $\beta_c$ increases marginally for progenitors with a larger $R_*$, the corresponding $\beta_h$ reaches relativistic velocities earlier due to less dense surrounding stellar medium.

Figure~\ref{comp_pressure} shows the time evolution of pressure for the external medium, the jet and the cocoon for a magnetized outflow propagating in the stellar medium. The results are shown for the same $(B_{\rm dip},P_i)$ configurations and density profiles as in Figure~\ref{comp_velocity}. 
As expected, for a given stellar density profile and jet opening angle, $P_j$ and $P_c$ are both higher for the outflows that are more energetic i.e. with larger $B_{\rm dip}$ and smaller $P_i$. For $B_{\rm dip} \sim 10^{15}-10^{16}\, {\rm G}$ and $P_i \sim 1-2\ {\rm ms}$, $P_j$ generally exceeds $P_c$ throughout the PNS spin-down evolution suggesting that the jet remains uncollimated prior to its breakout. The jet pressure increases with the initial opening angle as $P_j \sim L_j/r_j^2 \sim \theta_j^{-2}$ once $\sigma_0 \gtrsim 1$ and $L_j$ becomes roughly constant after few seconds (see Figure~\ref{sigma0_Edot_t}). Although both $P_c$ and $P_{\rm ext}$ are independent of $\theta_j \sim 5-20^{\circ}$, the former tends to be slightly lower for the BSG and RSG progenitors that have a larger $R_*$ (and therefore smaller $P_{\rm ext}$) in comparison to their WR counterparts. Consequently, it is easier for the surrounding medium to collimate the magnetized jets that have a wider $\theta_j$, and especially, in the WR progenitors.

\label{lastpage}

\end{document}